\documentclass{article}
\newcommand\rcsinfo{ \$Source: /home/axel/paper/spec/RCS/spec.tex,v $ $
  -- \$Revision: 1.53 $ $ -- \$Date: 2005/05/26 09:10:34 $ $ }

\usepackage[sort]{natbib}
\bibpunct{(}{)}{,}{a}{,}{,}

\newcommand{\var}{\mathop{\mathrm{var}}}
\newcommand{\std}{\mathop{\mathrm{std}}}
\newcommand{\cm}[1]{({\small \sf #1})} 
\usepackage{amsfonts}
\usepackage{amsmath} 
\usepackage{amssymb}
\usepackage{url} 
\usepackage{graphicx}
\usepackage[hypertex]{hyperref}
\newlength{\imgwidth}

\usepackage{version}
\includeversion{onlysubmission}\excludeversion{onlypreprint}
\excludeversion{motivation}

\usepackage{fullpage}

\def\ifUnDefinedCs#1{\expandafter\ifx\csname#1\endcsname\relax}
\ifUnDefinedCs{FreezeFont} \includeversion{nowordcount} \else
\excludeversion{nowordcount}
\renewcommand\citet[2]{\ignorespaces}
\renewcommand\cm[1]{\ignorespaces}
\renewcommand{\maketitle}{\ignorespaces} 
\fi
\renewcommand{\cm}[1]{\ignorespaces}
\title{Some Properties of the Speciation Model\\ for Food-Web
  Structure --- \\ Mechanisms for Degree Distributions and Intervality}

\author{A. G. Rossberg$^\ast$, H. Matsuda, T. Amemiya, K. Itoh\\
  \normalsize{Yokohama National University, Graduate School of Environment}\\
  \normalsize{and Information Sciences, Yokohama 240-8501, Japan}\\
  \normalsize{$^\ast$Corresponding author. Tel.:
    +81-45-339-4369, FAX: +81-45-339-4353}\\
\small{E-mail addresses: rossberg@ynu.ac.jp (A.G.R.),
  matsuda2@ynu.ac.jp  (H.M.),
 }\\\small{amemiyat@ynu.ac.jp (T.A.), itohkimi@ynu.ac.jp (K.I.)}}
\begin{onlypreprint}
   \date{\small \rcsinfo}
\end{onlypreprint}
\begin{onlysubmission}
  \date{\today}
\end{onlysubmission}

\begin{document}

\begin{nowordcount}
  \maketitle
\end{nowordcount}

\thispagestyle{headings}

\begin{abstract}
  \cm{We understand everything, but will the reader?  Will the
    referee? }
  
  We present a mathematical analysis of the speciation model for
  food-web structure, which had in previous work been shown to yield a
  good description of empirical data of food-web topology.  The degree
  distributions of the network are derived.  Properties of the
  speciation model are compared to those of other models that
  successfully describe empirical data.  It is argued that the
  speciation model unifies the underlying ideas of previous theories.
  In particular, it offers a mechanistic explanation for the success
  of the niche model of Williams and Martinez and the frequent
  observation of intervality in empirical food webs.
\end{abstract}

\newpage{}

\cm{Everything that looks like this is a comment and will not appear
  in the submitted version.}

\tableofcontents{}

\section{Introduction}
\label{sec:introduction}

The theoretical study of the topology of food webs, the networks
formed by the trophic interactions in ecological communities, has led
to increasingly precise descriptions of the empirically observed
structures.  In the early work of
\citet{cohen78:_food_webs_niche_space},
\citet{briand87:_envir_chain_length}, \citet{sugihara92:_niche} and
others, several simple food-web models had been investigated.  The
\emph{cascade model} \citep{cohen90:_commun_food_webs} was identified
as a description that reproduced the available data particularly well.
In the cascade model a food web consists of a fixed number $S$ of
species, and each species consumes any species which precedes it in a
given linear ordering with a fixed probability $C_0$.  The analysis of
this model led to several predictions
\citep{cohen90:_commun_food_webs} which inspired a more systematic and
accurate collection of food-web data by empiricists
\citep[e.g.,][]{polis91:_compl_des_web,martinez91:_artif_attr,hall91:_food_rich_web,havens92:_scale_webs}.

Based on the new data, \citet{williams00:_simpl} showed that their
\emph{niche model} was a significant improvement.  In this model,
species are ordered according to their niche value $n$ that is chosen
randomly from the interval $[0,1]$.  To determine the diet of a
species, an interval of random width $\le n$ is drawn with even
distribution from within\footnote{The original description of the
  model \citep{williams00:_simpl} is inaccurate at this point.}
$[0,1]$, restricted by the condition that at least half of the
interval is located below the niche value $n$ of this species.  Its
diet then consists of all species with a niche value in this interval.

A mathematical analysis by \citet{camacho02:_analytic_food_webs}
revealed the importance of the specific rule for determining the width
of the feeding intervals: by choosing it from an approximately
exponential distribution, the resulting food webs show a distribution
of generality (the number of a species' resources) which is strongly
skewed towards low values, in good accordance with observations
\citep{camacho02:_robus_patter_food_web_struc,%
stouffer05:_quant_patterns_webs}.

By construction, the niche model also reproduces a property called
\emph{intervality} \citep{cohen78:_food_webs_niche_space}: Species can
be ordered on a line in such a way that the diet of each consumer is a
contiguous set.  Intervality is surprisingly often found in small webs
\citep{cohen78:_food_webs_niche_space}.  Larger webs exhibit it to
some degree \citep{cohen90:_commun_food_webs,cattin04:_phylog}.
\citet{cattin04:_phylog} argued that intervality can be a consequence
of the fact that similar, evolutionary related species consume similar
resources.  They proposed the \emph{nested hierarchy model}, a
modification of the niche model which incorporates this idea and
better accounts for the observed degree of intervality.

Apart from these mostly descriptive models of food-web topology there
have also been several attempts to explain the structure of food webs
by the interaction of population dynamical and evolutionary mechanisms
\citep[e.g.,][]{caldarelli98:_model_multispec_commun,%
drossel01:_influen_predat_prey_popul_dynam,%
yoshida03:_evolut_web_sys,%
tokita03:_emerg_complex_stab_net}.  Characteristic for most of these
models is their high computational complexity, which makes their
quantitative statistical validation difficult.  Therefore it can be
advantageous to consider first explanatory models that are explicit in
terms of either population dynamics
\citep[e.g.,][]{pimm84:_compl_stab,montoya03:_topol_webs_real_to_assemb}
or evolutionary mechanisms
\citep[e.g.,][]{amaral99:_envir_chang_coext_patter_fossil_recor,drossel98:_extin_event_species_lifet_simpl_ecolog_model,camacho00:_extin}
alone.

The recently proposed \emph{speciation model} \citep{rossberg05:_web}
is of the purely evolutionary type.  It combines mechanisms
corresponding to speciations and extinctions with simple assumptions
regarding the evolutionary inheritance of trophic links.  In spirit,
the model is similar to the duplication-divergence model of proteome
evolution by \citet{vazquez03:_model_protein_net} or the related model
by \cite{pastor-satorras03:_evolv_prot_net}, even though in the
speciation model directed links and the possibility of extinctions
complicate the situation.  

Furthermore, the speciation model takes the tendency of food webs to
respect a ``pecking order'', as it is ideally realized in the cascade
model, into account.  It is currently unclear if the dominating mechanism
imposing this ordering of species is the physical advantage that larger
predators have over smaller prey, energy conservation and dissipation,
or some other constraint.  The idea that the pecking order is
essentially an ordering by body size has often been discussed
\citep{cattin04:_phylog,warren87:_pre_pry_triang,cohen93:_body_size,
  memmott00:_predat_size_web}.  The speciation model makes this
hypothesis explicit by postulating an allometric relationship between
body sizes and evolution rates.

The speciation model has been validated by a systematic statistical
analysis based on a comparison of twelve model properties---such as
the average chain length, the fraction of top predators, the degree of
intervality, or the clustering coefficient---with empirical data
\citep{rossberg05:_web}.  These numerical results suggest that the
speciation model reproduces observed food-web properties even better
than the niche model or the nested hierarchy model.
The aim of the current work is to present some analytic results that
allow insights into how important food web properties derive from the
model specifications.  After stating the model definition in
Sec.~\ref{sec:speciation_model}, the steady-state distribution of the
number of species $S$ and the expectation value of the directed
connectance $C$ (sometimes referred to as the food-web
``complexity'') are derived in Sec.~\ref{sec:analysis}.  These
quantities are important because they are used as control parameters
in other models.  Section~\ref{sec:analysis} also contains a
characterization of the species pool in terms of evolutionary
``clades'' which invites a comparison with empirical data.
Section~\ref{sec:distributions} is devoted to a characterization of
the model in terms of the distributions of generality and
vulnerability (the number of a species' consumers).  Based on these
results, the speciation model is compared with \cm{to} the cascade
model, the niche model, and the nested hierarchy model in
Sec.~\ref{sec:other-models}; common properties and differences are
pointed out.  Two variants of the speciation model, which leave the
analytic properties derived below unchanged, are introduced in
Sec.~\ref{sec:variants}.  A discussion and interpretation of the
results is provided in Sec.~\ref{sec:conclusion}.

\section{Definition of the speciation model}
\label{sec:speciation_model}

This section restates the definition of the speciation model given
elsewhere \citep{rossberg05:_web}, since it will be the starting point
for the subsequent analysis.  For a motivation of the model and a
discussion of design decisions we refer to the original work.
The speciation model describes an abstract species pool, the set of
trophic links between the species, and the evolution of both.  The
model is described in terms of a stochastic process characterized by
the parameters $r_1$, $r_+$, $r_-$, $R$, $D$, $\lambda$, $C_0$, and
$\beta$. 

\subsection{The evolution of the species pool}

Each species $i$ in the pool is associated with a \emph{speed
  parameter} $s_i$ in the range $[0,R]$.  The speed parameter
characterizes the evolution rate of a species and is thought to be
inversely correlated with the logarithm of the species' body mass by
an allometric law (see \citet{rossberg05:_web} for discussion).  In
any infinitesimal time interval $[t,t+dt]$ three kinds of events can
occur: \emph{adaptations} of foreign species to the habitat (i.e.
invasions on an evolutionary time scale), \emph{extinctions}, and
\emph{speciations}.  The probability for the adaptation of a new
species $k$ with speed parameter in the infinitesimal range
$s_k\in[s,s+ds]$ is $r_1 \exp(s)\, ds\, dt$.  When a new species is
adapting to the habitat, it is added to the species pool.  The
probability that some species $i$ of the species pool becomes extinct
is $r_- \exp(s_i) \, dt$.  When a species becomes extinct, it is
removed from the species pool.  Finally, the probability that some
species $i$ from the species pool speciates is $r_+ \exp(s_i) \, dt$.
When $i$ speciates, a new species $j$ with speed parameter
$s_j=s_i+\delta$ is added to the species pool, where $\delta$ is a
zero-mean Gaussian random variable with
$\mathop{\mathrm{var}}\delta=D$.  If $s_i+\delta$ exceeds the range
$[0,R]$, $s_j=-(s_i+\delta)$ or $s_j=2 R-(s_i+\delta)$ are used
instead (reflecting boundaries).  The probabilities for any of these
events to occur are independent.

\subsection{The evolution of the food web}

The food web is described by a connectivity (or adjacency) matrix
$(m_{ij})$, with connectivity values $m_{ij}=1$ when $j$ eats $i$ and
$m_{ij}=0$ otherwise.  \emph{Possible consumers} $l$ of species $i$
are defined as species with $s_l<s_i+\lambda\,R$, \emph{possible
  resources} $h$ as those with $s_h>s_i-\lambda\,R$.  The connectivity
$m_{ij}$ can be $1$ only when $i$ is a possible resource of $j$.  The
connectivity of a new species adapting to the habitat to all possible
consumers and resources is set to $1$ with probability $C_0$ and to
$0$ otherwise.  Upon speciation, the connectivity values of the
decedent species $j$ to possible consumers and resources are copied
from the corresponding connectivity values of the parent species $i$
with probability $1-\beta$ (i.e., links break with probability
$\beta$).  The connectivity values to all possible resources and
consumers of $j$ which have not been copied are set to $1$ with
probability $C_0$ and to $0$ otherwise.

\subsection{Typical parameters}

\begin{table}[btp]
  \centering
\begin{tabular}{lrrrrrrr}
  Food web: &BB&Sk&Co&Ch&SM&Yth&LR\\
  \hline
  model parameters: \\
  $r_+$ ($=\rho$)& 0.914&      0.934&    0.961&     0.959&     0.801&     0.949&       0.991\\
  $r_1$&           0.17&          0.21&    0.13&     0.21&     0.92&     0.67&      0.13\\
  $\lambda$&       0.12&      0.082&   0.006&   0.25&     0&            0.001&   0.025\\
  $C_0$&           0.37&      0.53&    0.58&     0.064&    0.23&     0.081&     0.16\\
  $\beta$&         0.059&      0.012&    0.014&     0.029&     0.034&
  0.040&      0.0063\\
  \hline
  derived quantities: \\
  web size (before lumping) $\left<S\right>$& 18.2&       29.0&     31.4&      47.9&      42.7&      122.0&       137.4\\
  $\var S/\left<S\right>^2$& 0.64&      0.53&    0.81&     0.51&     0.12&     0.16&      0.81\\
  clade size $\left<n\right>$: Eq.~(\ref{species-per-clade})& 4.3&       5.2&     7.6&      7.3&      2.5&      6.3&       23.5\\
  number of clades $\left<c\right>$: Eq.~(\ref{clade-in-web}) & 4.2&       5.5&     4.2&      6.6&      17.1&      19.5&       5.8\\
  clade lifetime in gen.: $-\ln(1-\rho)$& 2.5&       2.7&     3.2&      3.2&      1.6&      3.0&       4.7\\
  clades in diet: Eq.~(\ref{clade-in-diet}), $\Lambda=R$& 2.3 & 3.2& 2.8& 0.7& 4.5 &2.8&  1.5\\
  diet breakout: Eq.~(\ref{breakout})& 0.44&      0.16&    0.31&       0.41&     0.12&     0.43&      0.44\\
\end{tabular}
\caption{Maximum-likelihood model parameters for the speciation model
  obtained for seven empirical food webs and quantities derived
  thereof.   The abbreviations stand for BB: 
  Bridge Brook Lake \citep{havens92:_scale_webs}, Sk:
  Skipwith Pond \citep{warren89:_spatial_freshw_web}, Co: Coachella
  Desert \citep{polis91:_compl_des_web}, Ch: Chesapeake Bay
  \citep{baird89:_chesap_bay}, SM: St.~Martin Island
  \citep{goldwasser93:_const_carib_web}, Yth: Ythan Estuary
  \citep{hall91:_food_rich_web}, LR: Little Rock Lake 
  \citep{martinez91:_artif_attr}. }
\label{tab:parameters}
\end{table}

In our previous study \citep{rossberg05:_web} the predictions of the
speciation model were compared to empirical data, and maximum
likelihood fits of the model to empirical data sets for fixed $R=\ln
10^4$, $D=0.0025$, $r_-=1$ were computed.  For brevity we refer to
these parameter sets as ``typical values'' hereafter.  For the
convenience of the reader the fitted values are listed in
Table~\ref{tab:parameters} together with some derived expressions
relevant for the calculations below.

\section{Basic statistical properties of the model steady state}
\label{sec:analysis}

The number $S$ of species in a food web and the number $L$ of trophic
links connecting them belong to the simplest quantities used to
characterize food webs.  Often $L$ is expressed in terms of the
directed connectance $C=L/S^2$ or related quantities.  In what
follows, the steady-state distribution of $S$ and the expectation
value of $C$ for the speciation model are derived.
For these calculations, it is helpful to imagine the species pool as
being divided into clades.  Following
\cite{yoshida02:_long_living_fossils,yoshida03:_evolut_web_sys}, a
\emph{clade} is here defined as the group of all currently existing
descendant species of a \emph{founder species} that entered the
species pool through an adaptation process, in close correspondence with
the standard phylogenetic notion.  When $D$ is sufficiently small, the
speed parameter $s$ is approximately the same for all species in a
clades, and the ranges of $s$ covered by different clades do not
overlap.  We can then divide the $s$ axis into small intervals
$[s,s+\Delta s]$, and account for the number of species in each
interval separately.  The absence of overlap between clades is used
only as a trick to simplify accounting.  The final results do not
depend on this assumption.  The condition that the spread of $s$
within clades is small will be made more precise in the detailed
discussion of the clades in Section~\ref{sec:clades} below.

\subsection{The steady-state distribution of the species number $S$}
\label{sec:S}

In order to obtain the steady-state distribution of the total number
of species, consider first only a small interval $[s,s+\Delta s]$ on
the speed-parameter axis.
The master equation for the probability distribution $p_n$ of the
number $n$ of species in the interval  is given by
\begin{align}
  \label{probability_balance}
  \frac{d p_n}{dt}=&j_{n-1,n}-j_{n,n+1}\\
  \intertext{for $n\ge 1$ and}
  \label{dp0}
  \frac{d p_0}{dt}=&j_{0,1},
\end{align}
with the probability current $j_{n,n+1}$, resulting from the
balance of processes incrementing and decrementing~$n$, given by
\begin{align}
  \label{jnnp}
  j_{n,n+1}=& e^s
  \left[
    (n\,r_+ + r_1 \Delta s) p_n - (n+1) r_- p_{n+1} 
  \right].
\end{align}
The possibility of speciations that cross the boundaries of the range
$[s,s+\Delta s]$ is ignored here, because the corresponding
corrections would cancel out when summing up the $n$ values from
different intervals below.  The reflecting boundary conditions at the
endpoints of the full $s$-range $[0,R]$ ensure that~(\ref{jnnp}) holds
also for the intervals adjacent to the endpoints.

For the steady state $j_{n,n+1}=0$ one gets 
\begin{align}
  \label{p1}
   p_1=&\frac{r_1}{r_-}\,p_0\,\Delta s\\
   \intertext{and for $n\ge 1$ the recursive relation}
   \label{pn}
   p_{n+1}=&\frac{n\,r_+}{ (n+1) r_-} p_n +\mathcal{O}(\Delta s),\\
   \intertext{which is solved by}
   \label{pn2}
   p_{n}=&\frac{1}{n}
   \left(
     \frac{r_+}{r_-}
   \right)^n \frac{r_1 p_0}{r_+} \Delta s + \mathcal{O}(\Delta s^2).
\end{align}
With the abbreviations $\rho=r_+/r_-$ and $\kappa=r_1/r_+$, the
corresponding moment generating function is
\begin{align}
  \label{momds}
  m(z)=
  \left<
    z^n
  \right>=p_0\,
  \left[
    1- \kappa\,\Delta s\,\ln
      \left(
      1-\rho z
      \right) \right]+\mathcal{O}(\Delta s^2),
\end{align}
with
\begin{align}
  \label{p0}
  p_0=1+\kappa\, \Delta s\ln(1-\rho)+\mathcal{O}(\Delta s^2)
\end{align}
given by the normalization condition $m(1)=1$.  From $m(z)$ one
obtains the cumulant generating function
\begin{align}
  \label{kumds}
  \begin{split}
    k(z)=\ln m(z)=&\ln p_0 - 
    \kappa\,\Delta s\ln\left( 1-\rho z \right)+\mathcal{O}(\Delta s^2)\\
    =&\kappa\,\Delta s\,
      \ln\!\left( \frac{1-\rho}{1-\rho z}
    \right)
    +\mathcal{O}(\Delta s^2).
    \end{split}
\end{align}
Cumulant generating functions of this form and the corresponding
distributions are discussed in Appendix~\ref{sec:general}.  For
example, by Eq.~(\ref{meanAB}), the density of species along the
speed-parameter line is
\begin{align}
  \label{density}
  \lim_{\Delta s\to 0}\frac{ \left< n \right>}{\Delta s}=\frac{\kappa
    \rho}{1-\rho}=\frac{r_1}{r_--r_+}.
\end{align}

\begin{figure}[tbp]
  \centering
  \includegraphics[width=\imgwidth,keepaspectratio,clip]{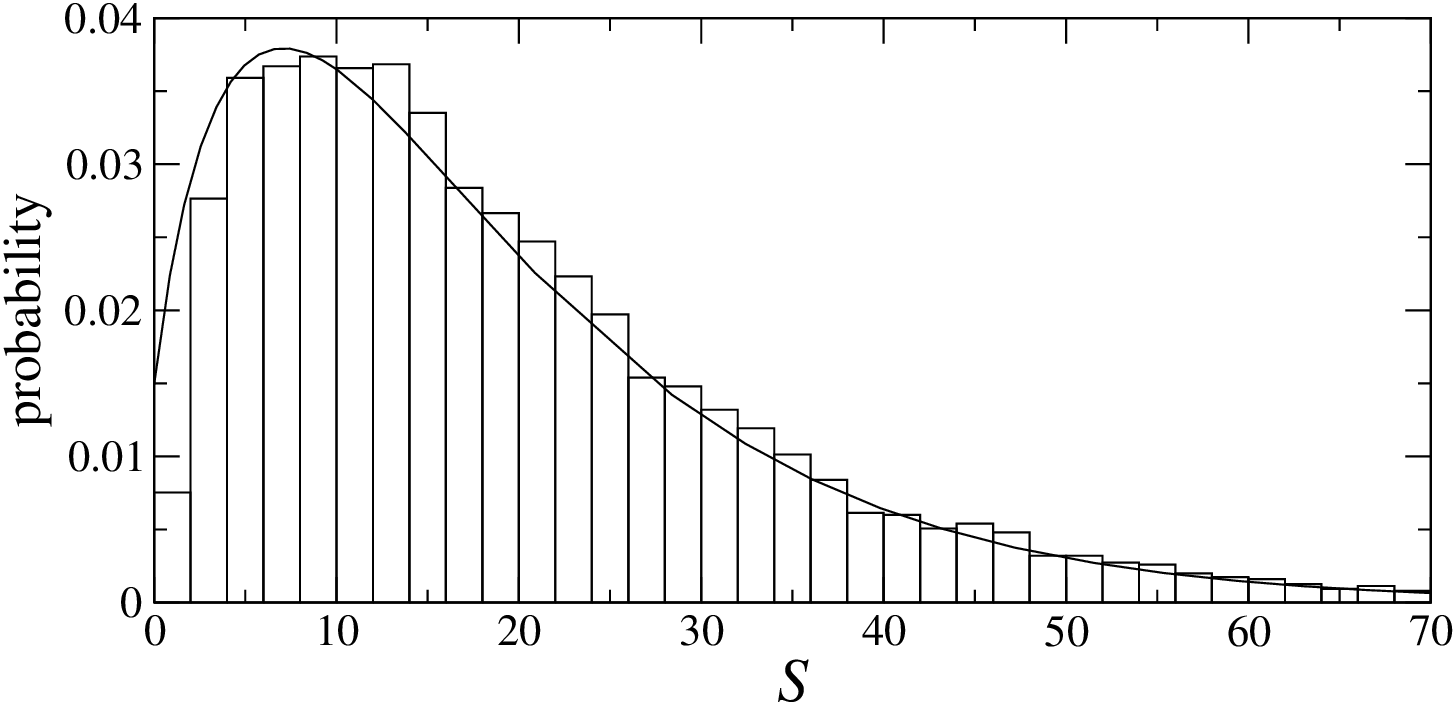}
  \caption{Typical steady-state distribution of the number of species
    $S$.  The solid line is $P(\kappa R,\rho;S)$ as defined by
    Eq.~(\ref{ABdist}); the histogram was obtained by direct
    simulations.  Parameters correspond to Bridge Brook Lake
    (Tab.~\ref{tab:parameters}).}
  \label{fig:S-distribution}
\end{figure}

The cumulant generating function of the sum of independent random
variables is the sum of their cumulant generating functions.  Thus,
the cumulant generating function for the total number of species $S$
can be obtained by dividing the range $[0,R]$ into small intervals of
width $\Delta s$ and summing the contributions.  With $\Delta s\to 0$
corrections $\mathcal{O}(\Delta s^2)$ become negligible and the
summation goes over into an integration:
\begin{align}
  \label{kums}
  \sum \frac{k(z)}{\Delta s} \Delta s +\mathcal{O}(\Delta s^2) \to
  \int_0^R \kappa
      \ln\left( \frac{1-\rho}{1-\rho z}
    \right)\,ds=\kappa R\ln
  \left(
    \frac{1-\rho}{1-\rho z}
  \right).
\end{align}
This is again of the general form Eq.~(\ref{generalK}) discussed in
Appendix~\ref{sec:general}.  Hence, the steady-state distribution of
the species number $S$ is $P(\kappa R,\rho;S)$ as defined by
Eq.~(\ref{ABdist}).  Figure~\ref{fig:S-distribution} shows a typical
distribution and corresponding simulation results.  The curves agree
well.  Only the probability for $S$ near zero seems to be
overestimated by the theory.  By Eq.~(\ref{meanAB}), the mean number
of species is
\begin{align}
  \label{meanS}
  \left<
    S
  \right>=\frac{\kappa R \rho}{1-\rho}
\end{align}
and by Eq.~(\ref{varAB}) the relative variance $(\var
S)/\left<S\right>^2=1/\kappa R \rho$. Typical relative variances
(Tab.~\ref{tab:parameters}) can become of the order unity.  Thus, in
the model, $S$ fluctuates strongly on evolutionary time scales.

\subsection{Basic properties of clades}
\label{sec:clades}

The division of $S$ into clades can be made more explicit.  For
example, the distribution of the number $n$ of species in a single
clade is given by Eq.~(\ref{pn2}) conditional to $n\ge 1$:
\begin{align}
  \label{clade-size-distribution}
  p_n=-\frac{\rho^n}{n \ln(1-\rho)}
\end{align}
Thus, the mean number of species per clade is 
\begin{align}
  \label{species-per-clade}
  \left<
    n
  \right>=\sum_n n\,p_n=-\frac{\rho}{(1-\rho)\,\ln(1-\rho)}.
\end{align}
Further, the expectation value of the number of clades $c$ in the food
web can be estimated as
\begin{align}
  \label{clade-in-web}
  \left<
    c
  \right>=\frac{
    \left<
      S
    \right>}{
    \left<
      n
    \right>}=-\kappa\, R \,\ln(1-\rho).
\end{align}
(An exact calculation yields the same result.)  Since appearances and
extinctions of clades are statistically independent, the number of
clades is Poisson distributed.  For typical values of $\left< n
\right>$ and $\left< c \right>$ see Tab.~\ref{tab:parameters}.

To obtain the average lifetime $\tau_c$ of a clade founded by a
species with speed parameter $s$, notice that the probability that a
clade exists in the interval $[s,s+\Delta s]$ is $1-p_0$ with $p_0$
given by~Eq.~(\ref{p0}).  On the other hand, new clades are founded at
a rate $r_1 \sigma\, \Delta s$ with $\sigma=\exp(s)$.  The fraction of
time when there is a clade in the interval is thus $\tau_c\, r_1
\sigma\, \Delta s$. (Note that in the limit $\Delta s\to 0$ there is no overlap
in the clade lifetimes.)  Thus
\begin{align}
  \label{tau-clade}
  \tau_c=\lim_{\Delta s\to 0} \,\frac{1-p_0}{r_1\sigma\Delta
    s}=-\frac{\ln(1-\rho)}{r_+\sigma}.
\end{align}
The time that it takes for the system to reach the steady state can be
estimated by the lifetime of the slowest clade, i.e., by
Eq.~(\ref{tau-clade}) with $\sigma=\exp(0)=1$.  This quantity is
important for model simulations.  For a detailed discussion of the
dynamics of the birth/death process relevant here, including the clade
lifetime distribution, see the book of \cite{bailey64:_stoch_proces}.

The typical number of evolutionary ``generations'' that a clade exists
is $\tau_c/\text{(generation time)}=\tau_c r_+ \sigma=-\ln(1-\rho)$
(see Tab.~\ref{tab:parameters} for typical values).  Since in each
generation the variance of the distribution of $s$ over a clade
increases by $D$, the width of a clade on the speed-parameter
line is of the order
\begin{align}
  \label{cladwidth}
  \std s\approx\sqrt{- D \ln(1-\rho)}.
\end{align}
The assumption made above that all
members of a clade have approximately the same $s$ is justified when
$s\ll1$. 

\subsection{The expected directed connectance}
\label{sec:links}

A food-web property that has found much attention in both empirical
and theoretical research is the connectance, for example measured in
terms of the directed connectance $C=L/S^2$
\citep{martinez91:_artif_attr} with $L$ denoting the total number of
trophic links.  To compute the expectation value of this quantity,
note that from all $S^2$ topologically possible links only some are
allometrically possible in the model, namely those from consumers $i$
to their possible resources $h$ with $s_h > s_i-\lambda R$ (s.\ 
Sec.~\ref{sec:speciation_model}).  A fraction $(1-\lambda)^2/2$ of the
$s_h$-$s_i$ plane is forbidden.  By construction, exactly a fraction
$C_0$ of all allometrically possible links is realized on the average
in the model.  Thus, as a simple estimate one gets $S^2
[1-(1-\lambda)^2/2]=S^2 (1+2 \lambda-\lambda^2)/2$ allometrically
possible links and
\begin{align}
  \label{simple-C}
  C\approx C_0 (1+2\lambda-\lambda^2)/2.
\end{align}
The exact value differs due to subtle correlations stemming from
intra-clade links.  As an example, we derive $C$ for the case that the
typical intra-clade spread of $s$ given by Eq.~(\ref{cladwidth}) is
much smaller than $\lambda R$, so that all intra-clade links are
allometrically possible.  As in Sec.~\ref{sec:S}, we divide the $s$ axis
into small intervals of width $\Delta s$, and do again as if each
clade was located in its own interval.  Let the $p$-th interval range
from $s_p$ to $s_p+\Delta s=s_{p+1}$ and denote the number of species
it contains by $n_p$.  We first compute the expected number of
allometrically possible links conditional to fixed $S$
\begin{align}
  \label{L}
  \left<
    L_\text{al}|S
  \right>=\mathop{\sum_{p,q}}_{s_p>s_q-\lambda R}
  \left<
    n_p\,n_q|S
  \right>=\mathop{\sum_{p\ne q}}_{s_p>s_q-\lambda R}
  \left<
    n_p\,n_q|S
  \right>+\sum_{p}
  \left<
    n_p^2|S
  \right>.
\end{align}
Consider the last term first.  The distribution $p_n$ of $n_p$ is
given by Eqs.~(\ref{pn2},\ref{p0}).  Since clades appear and disappear
independently, the probability that there are $S-n_p$ species outside
the $p$-th interval is, just as for the total number of species,
$P(\kappa R,\rho;S-n_p)$, defined by Eq.(\ref{gen-dist}) to lowest
order in $\Delta s$.  The probability for a particular pair $(n_p,S)$
is therefore $p_{n_p}\,P(\kappa R,\rho;S-n_p)$.  This can be used to
calculate the probability $p(n_p|S)$ of $n_p$ conditional to $S$ in
the usual way, giving
\begin{align}
  \label{np2}
  \left<
    n_p^2|S
  \right>
  =\sum_{n=0}^S n^2\,p(n|S)
  =\sum_{n=0}^S n^2\, \frac{p_n P(\kappa
      R,\rho;S-n)}{P(\kappa R,\rho;S)}
  =
  \frac{S\,(\kappa
    R+S)
  }{R\,(1+\kappa R)} \Delta s+\mathcal{O}(\Delta s^2).
\end{align}
The dependence on $\rho$ drops out.  By a similar argument one obtains
to lowest order in $\Delta s$
\begin{align}
  \label{npnq}
  \left<
    n_p n_q|S
  \right>
  =\sum_{m+n\le S} n\,m\,p(m,n|S)
  =
  \frac{\kappa (S-1) S}{R\,(1+\kappa R)} \Delta s^2.
\end{align}
Inserting both results into~(\ref{L}) and taking the limit $\Delta
s\to 0$ yields
\begin{align}
  \label{L2}
  \left<
    L_\text{al}|S
  \right>=&S\,\frac{S+\kappa R\,
    \left[
    1+\frac{1}{2}(1+2\lambda-\lambda^2)\,(S-1)
    \right]}{1+\kappa R}\\
  \label{L2approx}
  = &S^2 \,\left[ \frac{1+\kappa
    R\frac{1}{2}(1+2\lambda-\lambda^2)}{1+\kappa R}+\mathcal{O}
  \left(
  \frac{\kappa R}{S}
  \right)\right].
\end{align}
Expression~(\ref{L2approx}) is often a good approximation
of~(\ref{L2}).  The expected directed connectance conditional to $S$
is $ \left< C|S \right>=C_0 \left< L_\text{al}|S \right>/S^2$.
Dropping the undefined case $S=0$, the expected connectance for
freely fluctuating $S$ can be evaluated as
\begin{align}
  \label{C}
  \left<C\right> = C_0 
  \left[
    1-P(\kappa R,\rho;0)
  \right]^{-1}\,
  \sum_{S=1}^\infty \frac{
    \left<
      L_\text{al}|S
    \right>}{S^2}\,
  P(\kappa R,\rho;S),
\end{align}
either directly numerically or, for a (complicated) closed-form
expression, with the help of symbolic algebra software.  For the
parameters of Bridge Brook Lake (Tab.~\ref{tab:parameters}), for which
$\lambda R/\std s=14.7$, Eq.~(\ref{C}) yields $\left<C\right>=0.294$
while the simple estimate~(\ref{simple-C}) gives
$\left<C\right>=0.230$.  Simulations yield $\left<C\right>=0.286$.
The cases that $\lambda R>0$ is much smaller than the typical
intra-clade spread of $s$ and that $\lambda=0$ (no cannibalism) can be
handled by replacing $n^2$ in Eq.~(\ref{np2}) by $n(n+1)/2$ or
$n(n-1)/2$ respectively.  For both cases the approximation
$\left<C|S\right>= \frac{1}{2}\,C_0 [1+\kappa R (1+2
\lambda-\lambda^2)]/(1+\kappa R)+\mathcal{O}(S^{-1})$ holds.  For the
parameters of St.\ Martin Island ($\lambda=0$) this yields
$\left<C\right>=0.115$, while numerically $\left<C\right>=0.112$ is
obtained.

\section{The distributions of generality and vulnerability}
\label{sec:distributions}

In this Section, analytic approximations for the distributions of
generality $k$ (the number of resources of a consumer) and
vulnerability $m$ (the number of consumers of a resource) are derived.
When defining the direction of trophic links in the standard way from
the resource to the consumer, these are the distributions of the
in-degree and the out-degree of the food web, respectively.  Degree
distributions are often thought to belong to the major determinants of
the overall network topology.  Due to the inherent randomness of food
webs and their finite size, instances of degree distributions of
empirical or model webs are also random quantities. Nevertheless, they
contain information regarding the probability distributions of
generality $P_\text{gen}(k)$ and vulnerability $P_\text{vul}(m)$ in
the steady state.  Specifically, if $N(k)$ denotes the number of
species with generality $k$ in a web and the total number of species
is $S$, then $\left< N(k)/S \right>=P_\text{gen}(k)$ in the steady
state.  While this is trivial for fixed $S$, it is worth noting that
this relation is valid also when the value of $S$ fluctuates randomly
and when the generalities of individual species are strongly
correlated with each other and with $S$, as can be seen by a
straightforward calculation.
Below it is shown that the conditional probability
$P_\text{gen}(k|S)$, i.e. the conditional expectation value
$\left<N(k)/S|S \right>$, does in fact strongly depend on $S$.  For a
comparison with single instances of empirical distributions $N(k)/S$
the conditional distribution $P_\text{gen}(k|S)$ is therefore better
suited than $P_\text{gen}(k)$.  Similar considerations hold for the
vulnerabilities.  Thus, the conditional distributions are computed
below.

Following \cite{camacho02:_analytic_food_webs}, we consider the
distinguished limit of large food-web sizes $S$ and small
connectances $C$ while keeping the link density $Z:=L/S=CS$ fixed.
(Fixing $Z$ for asymptotic expansions is not meant to suggest that $Z$
is actually fixed for large food webs.)  For simplicity, we make use
of the hypothesis that resources typically evolve faster than their
consumers in the extreme form that resources evolve \emph{much} faster
than their consumers.  This corresponds to assuming a large spread of
time scales $R$ and a small loopiness $\lambda$.  Errors due to
intra-clade trophic links, which violate this hierarchy of timescales,
are small when the total number of clades~(\ref{clade-in-web}) is
large, due either to large $\kappa R$ or to small $1-\rho$.  We note
that in the case $\kappa R\gg 1$ the combined effect of these
assumptions would reduce the formula for the directed connectance
derived above to $\left<C\right>=C_0/2$, which shows that the
approximations employed here are much coarser than those used in the
forgoing Sections.  Nevertheless they retain the main effects that
determine the general forms of the degree distributions.

\subsection{Reduction to the dynamics of the actual resources}
\label{sec:actual-resources}

When most resource species evolve much faster than their consumers,
the distribution of generality for a given consumer can be
approximated by the steady-state generality distribution with the
consumer assumed fixed while its resources evolve.
We first show that, using a simple mean-field-type approximation, the
stochastic dynamics of the actual resources of the fixed consumer can
be separated from the dynamics of the possible resources which are not
actual resources (called \emph{spurned resources} below) in a
self-consistent way.

To derive the dynamics of the actual resources, consider a small
interval $[s,s+ds]$ in the range of possible resources.  Let
$\sigma=\exp(s)$.  The rate at which actual resource species in the
interval speciate in such a way that the descendant species remain actual
resources is $r_+^*\,\sigma$ with
\begin{align}
  \label{rpB}
  r_+^*=(1-\beta) r_+ + \beta\,C_0 r_+.
\end{align}
The first term accounts for trophic links that do not break in the
speciation, and the second term for trophic links that break but are
immediately reconnected.  The probability that a resource species
becomes extinct in a time interval of length $dt$ is simply $r_-^*\,
\sigma \,dt$ with
\begin{align}
  \label{rmB}
  r_-^*=r_-.
\end{align}
Finally, the consumer can acquire a novel resource species either by
an adaptation of a new species to the habitat or by a speciation of a
spurned resource in such a way that the decedent species becomes an
actual resource.  For the rate at which the latter event occurs, a
mean-field type approximation is employed: The number of spurned
resources $n^\circ$ in the speed-parameter range $[s,s+\Delta s]$ is
approximated by its expectation value $ \left< n^\circ \right>$.  The
rate at which a predator acquires novel resources (that did not
speciate from an existing resource species) in this range is then
given by $C_0 r_1^* \,\sigma\, \Delta s$ with
\begin{align}
  \label{r1B}
  r_1^*=r_1+  \frac{\beta\,
  \left<
    n^\circ
  \right> r_+}{\Delta s}.
\end{align}
The first term represents new adaptations, the second term mutations of
spurned species.  With this approximation, the expectation value for
the number $n^*$ of actual resources in the range $[s,s+\Delta s]$ can
be calculated as
\begin{align}
  \label{nB}
  \left<
    n^*
  \right>=\frac{C_0\,r_1^*}{r_-^*-r_+^*}\, \Delta s
\end{align}
by methods analogue to those used in Section~\ref{sec:S}.  Deviations
from this mean-field approximation occur because the expectation value
$\left< n^\circ \right>$ is correlated to $n^*$ by the breaking of
actual links, which occurs at a rate $\mathcal{O}(\beta)$.  Since the
contribution of $\left< n^\circ \right>$ to the dynamics of $n^*$ is
also of order $\mathcal{O}(\beta)$, the resulting error in the
distribution of $n^*$ is $\mathcal{O}(\beta^2)$.

For the dynamics of the number of spurned resource, a set of equations
corresponding to Eqs.~(\ref{rpB}-\ref{nB}) can be set up by replacing
$C_0\to 1-C_0$ and interchanging the indices $*$ and
$\circ$~(\ref{rpB}-\ref{nB}).  These equations can be used to
eliminate $\left< n^\circ \right>$ from Eq.~(\ref{r1B}), yielding
\begin{align}
  \label{r1Bnew}
  r_1^*=r_1+\frac{\beta\,(1-C_0)\, r_+}{r_- - r_+}\,r_1.
\end{align}

\subsection{The generality distribution for fluctuating $S$}
\label{sec:generality-free-S}

Analogous to the calculations of Section~\ref{sec:S}, the cumulant
generating function for the number of actual resources for a species
with speed parameter $s$ can now be obtained as
\begin{align}
  \label{cumulantk}
  K_\text{gen}(s,z)=C_0 \,\kappa^* \Lambda(s) \ln \left( \frac{1-\rho^*}{1-\rho^*
      z} \right),
\end{align}
where $\rho^*=r_+^*/r_-^*$ and $\kappa^*=r_1^*/r_+^*$ are given by
Eqs.~(\ref{rpB},\ref{rmB},\ref{r1Bnew}), and
$\Lambda(s)=\min[(1+\lambda)\,R-s,R]\approx R-s$ is the size of the
speed-parameter range of possible resources.  The corresponding
distribution function is
\begin{align}
  \label{Pgen}
  P_\text{gen}(s,k)=P(C_0\,\kappa^* \Lambda(s),\rho^*;k)
\end{align}
as defined by Eq.~(\ref{ABdist}).  In particular, the expected number of a
consumer's resources is
\begin{align}
  \label{averagem}
  \left< k \right>=\frac{C_0\,\kappa^* \Lambda(s) \rho^*}{1-\rho^*}=C_0
  \Lambda(s)\,\frac{r_1}{r_--r_+}.
\end{align}
Comparison with Eq.~(\ref{density}) shows that the mean-field
approximation preserves the model property that, on the average, a
fraction $C_0$ of all allometrically possible links is realized.  Just
as for the overall species pool, the diet of a consumer can be divided
into several clades, each descending from a single newly acquired
resource.  For example, the expected number of resource clades for a
consumer is
\begin{align}
  \label{clade-in-diet}
  -C_0\kappa^* \Lambda(s) \ln (1-\rho^*),
\end{align}
in analogy to Eq.~(\ref{clade-in-web}).  

\begin{figure}[tbp]
  \begin{center}
    \includegraphics[width=\imgwidth,keepaspectratio,clip]{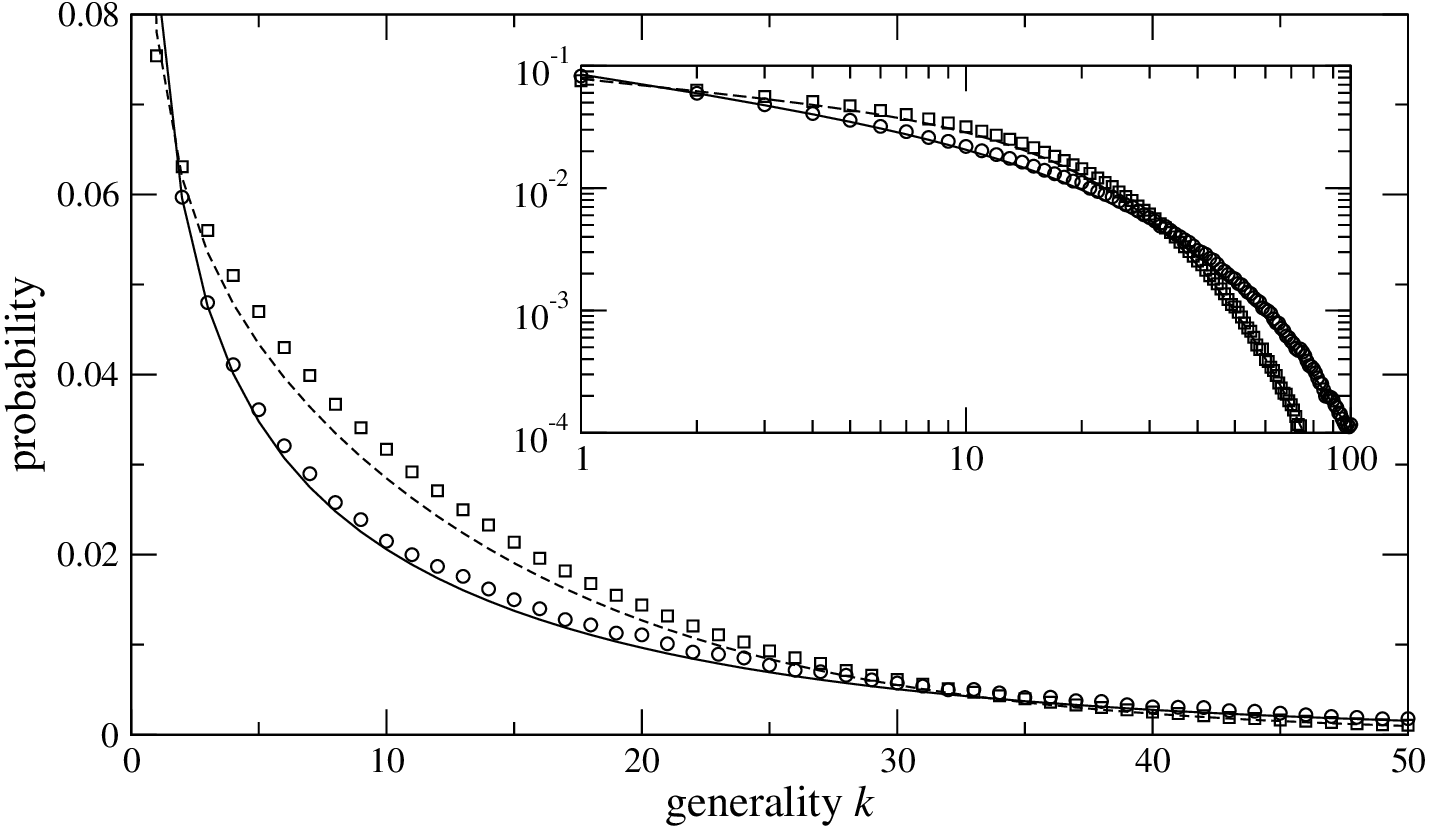}
  \end{center}
  \caption{Steady-state generality distributions for fluctuating
    species number, calculated from Eq.~(\ref{Mgen-overall}) with
    $\beta=0$ (solid) and $\beta=0.05$ (dashed) in comparison with
    direct numerical simulations (circles, squares).  The other
    parameters were $R=\ln 10^{20}$, $D=0.005$, $\rho=0.95$,
    $\kappa=10/R$, $C_0=0.1$, $\lambda=10^{-3}$ (no fitting).  The
    inset shows the same data on a double-logarithmic scales.}
  \label{fig:fluctuating-S-dists}
\end{figure}

Since, on the average, species are homogeneously distributed along
$s$, the probability distribution $P_\text{gen}(k)$ of the generality
of a species chosen arbitrarily from a food web is, to a good
approximation, the average of $P_\text{gen}(s,k)$ over $s$.
Analytically, this average is more easily calculated in terms of the
moment generating function $M_\text{gen}(s,z)=\sum_k
P_\text{gen}(s,k)\,z^k=\exp\,K_\text{gen}(s,z)$.  For the simple case
$\lambda=0$ one obtains
\begin{align}
  \label{Mgen-overall}
  M_\text{gen}(z)=\frac{1}{R}\int_0^R M_\text{gen}(s,z)\,ds=
  \frac{u-1}{\log u}\quad\text{with } u=
    \left(
      \frac{1-\rho^*}{1-\rho^*\,z}
    \right)^{\textstyle C_0 \kappa^* R}.
\end{align}
The generality distribution $P_\text{gen}(k)$ itself can be calculated
by a Taylor expansion of $M_\text{gen}(z)$ in $z$ or numerically from
the Fourier transformation of
$\mathop{\mathrm{Re}}\{M_\text{gen}(e^{i\phi})\}$.  A comparison with
direct numerical simulations shows that the condition that $R$ is
large is important for the numerical validity of
Eq.~(\ref{Mgen-overall}).  For example,
Fig.~\ref{fig:fluctuating-S-dists} shows analytic and numerical
results for $R=\ln 10^{20}$ in good agreement.

\subsection{The generality distribution for fixed $S$}
\label{sec:generality-fixed-S}

In order to compute the generality distribution $P_\text{gen}(k|S)$
conditional to fixed $S$, we start again from the distribution
$P_\text{gen}(s,k|S)$ for a consumer with speed parameter $s$.  In
order to simplify the calculations $\beta=0$ is assumed here.  Then
$\kappa^*=\kappa^\circ=\kappa$ and $\rho^*=\rho^\circ=\rho$.

For a given consumer, the species pool can be divided into three
subsets: (i) the actual resources of the consumer, (ii) the
allometrically possible but spurned resources, and (iii) the
allometrically forbidden resource (see Sec.~\ref{sec:links}).  For
small enough $D$, each clade is located in a single subset, and the
species distributions in the three subsets become independent.  We
first calculate the probability distribution for the number of species
in the union of the sets (ii) and (iii) for freely fluctuating $S$.
As above, denote the width of the range of allometrically possible
resources on the $s$ axis by $\Lambda$.  The distribution of the
species number in set (ii) can be obtained from Eq.~(\ref{cumulantk})
by substituting $C_0 \to 1-C_0$ and is therefore given by
$P((1-C_0)\kappa \Lambda,\rho;n)$ as defined in Eq.~(\ref{ABdist}).
The distribution of the number of species in (iii) can be obtained in
the same way as the distribution of the total number of species
(Sec.~\ref{sec:S}), just that the relevant range of $s$ is now
$R-\Lambda$, and not $R$.  Hence this distribution is given by
$P(\kappa \,(R-\Lambda),\rho;n)$.  The distribution of the number of
species in the union of these two sets is given by the convolution
\begin{align}
  \label{uni-dist}
  P_\text{union}(n)=P((1-C_0) \kappa
  \Lambda,\rho;n)*P(\kappa \,(R-\Lambda),\rho;n) =
  P(\kappa\,(R-C_0\Lambda),\rho;n).
\end{align}
The second equation is easily verified by comparing the corresponding
cumulant generating functions~(\ref{generalK}).

The number of species in set (i) is given by
$P_\text{gen}(k)=P(C_0\kappa\Lambda,\rho;k)$ as defined above.  Using the
known distribution $P(\kappa R,\rho;S)$ for $S$, the conditional
distribution of generality can be obtained as
\begin{align}
  \label{gen-dist}
  \begin{split}
    P_\text{gen}(k|S) =&\, \frac{P_\text{gen}(k) \,
      P_\text{union}(S-k) }{ P(\kappa R,\rho;S) }\\
    =&\, \frac{
      \Gamma(C_0 \kappa \Lambda + k)\,
      \Gamma(\kappa R)\,
      \Gamma(1 + S)\,
      \Gamma( \kappa\,(R-C_0\Lambda)-k+S)
   }{
     \Gamma(C_0\kappa\Lambda)\,
     \Gamma(1 + k)\, 
     \Gamma(\kappa\,(R-C_0 \Lambda))\,
     \Gamma(1 - k + S)\,
     \Gamma(\kappa  R + S)}.
\end{split}
\end{align}
Remarkably, just as the conditional expectations Eq.~(\ref{np2}) and
(\ref{npnq}), this result is independent of $\rho$.  The parameters
$S$ is playing a similar role instead (see below).
Equation~(\ref{gen-dist}) is now evaluated for large $S$.
Specifically, we assume (i) $S\gg\kappa R$, which is natural when $S$
is of the order of its expectation value $\kappa R \rho/(1-\rho)$ and
$1-\rho\ll 1$, (ii) $S \gg 1$, (iii) we restrict ourselves to values
of $k\ll S$, and (iv) in order to take the distinguished limit of
fixed link density, we set $C_0=Z_0/S$ with fixed $Z_0$.
Expanding the logarithm of Eq.~(\ref{gen-dist}) for large $S$ (e.g.,
using Stirling's formula) then gives
\begin{align}
  \label{large-S-expansion}
  \ln P_\text{gen}(k|S)=&\ln \frac{\kappa \Lambda Z_0}{k\,S}+
  \frac{
    -(\kappa R-1)\,k+\kappa L Z_0\,
    \left[
      \gamma+\psi_0(\kappa R)+\psi_0(k) -\ln S
    \right]
  }{S} + \cdots,
\end{align}
where $\gamma\approx 0.57$ is the Euler constant and
$\psi_0(x)=(d/dx)\ln\Gamma(x)$ the digamma function. 

A similar expansion can be obtained for a distribution of the form
$P(C_0 \kappa \Lambda,\tilde \rho;k)$ given by Eq.~(\ref{ABdist}),
when the parameter $\tilde \rho$ is assumed to behave such that
$S=b/(1-\tilde \rho)$ with fixed $b$ for large $S$, which is natural
in view of $ \left< S \right>\sim 1/(1-\rho)$.  One obtains
\begin{align}
  \label{Pgen-large-S}
  \ln P(C_0 \kappa \Lambda,\tilde \rho;k)=\ln \frac{\kappa \Lambda
  Z_0}{k\,S}+
\frac{
    - b\,k+\kappa L Z_0\,
    \left[\gamma+\ln b+
      \psi_0(k)-\ln S
    \right]
  }{S} + \cdots.
\end{align}
A comparison of the two expansions shows that 
\begin{align}
  \label{compare-dists}
  P_\text{gen}(s,k|S)\approx \mathcal{N}\, P(C_0 \kappa
  \Lambda(s),\tilde \rho;k),
\end{align}
where
\begin{align}
  \label{normalizator}
  \mathcal{N}=\mathcal{N}(s)=\exp
  \left\{
    C_0\kappa \Lambda(s)\,
    \left[
      \gamma_0(\kappa R)-\ln (\kappa R-1)
    \right]
  \right\}
\end{align}
and
\begin{align}
  \label{tilde-rho}
  \tilde\rho=1-\frac{\kappa R-1}{S}.
\end{align}
Hence, apart from the new parameters $\mathcal{N}$ and $\tilde \rho$,
the form of the generality distribution for fixed $S$ is approximately
the same as for fluctuating $S$.

The additional normalization factor $\mathcal{N}$ enters because $k$
can never exceed $S$, while $P(C_0 \kappa \Lambda,\tilde \rho;k)$ is
nonzero for all $k$.  When the expected number of consumers is much
smaller than $S$, i.e., for small connectances $C_0$, the value of
$\mathcal{N}$ approaches 1.  This can be seen by noting that
$\gamma_0(x)-\ln (x-1)= 1/(2x)+\mathcal{O}(x^{-2})$, so that we can
write $\mathcal{N}=\exp[\tilde C \Lambda/(2 R)]$ with $\tilde C\approx
C_0$.
The dependence on $S$ is fully contained in the new parameter $\tilde
\rho$.  Its relation to $\rho$ can be understood by substituting $S$
in Eq.~(\ref{tilde-rho}) by $\left<S\right>=\kappa R\rho/(1-\rho)$, which
leads to
\begin{align}
  \label{tilde-rho2}
  1-\tilde\rho=\frac{\kappa R-1}{\kappa R\rho}\,(1-\rho)\approx (1-\rho).
\end{align}
Of course, the forgoing interpretation of Eq.~(\ref{compare-dists})
makes sense only when $\kappa R>1$.  Yet, Eq.~(\ref{compare-dists}) is
numerically valid also when continued analytically to the region
$0<\kappa R \le 1$ where $\tilde \rho \ge 1$.

In Section~\ref{sec:actual-resources} it was shown that the effect of
a small, non-zero $\beta$ can be approximated by a renormalization of
the coefficients $\kappa$ and $\rho$.  Equation~(\ref{compare-dists})
shows that for $\beta=0$ the effect of fixing $S$ is also essentially
a renormalization of $\rho$.  Even though the generality distribution
for fixed $S$ and non-zero $\beta$ is difficult to compute
analytically, it is reasonable to assume that this too can be
approximated by an expression of the form~(\ref{compare-dists}) with
an appropriate pair of parameters $\tilde \kappa$ and $\tilde \rho$.

\begin{figure}[tbp]
  \centering
  \includegraphics[width=\imgwidth,keepaspectratio,clip]{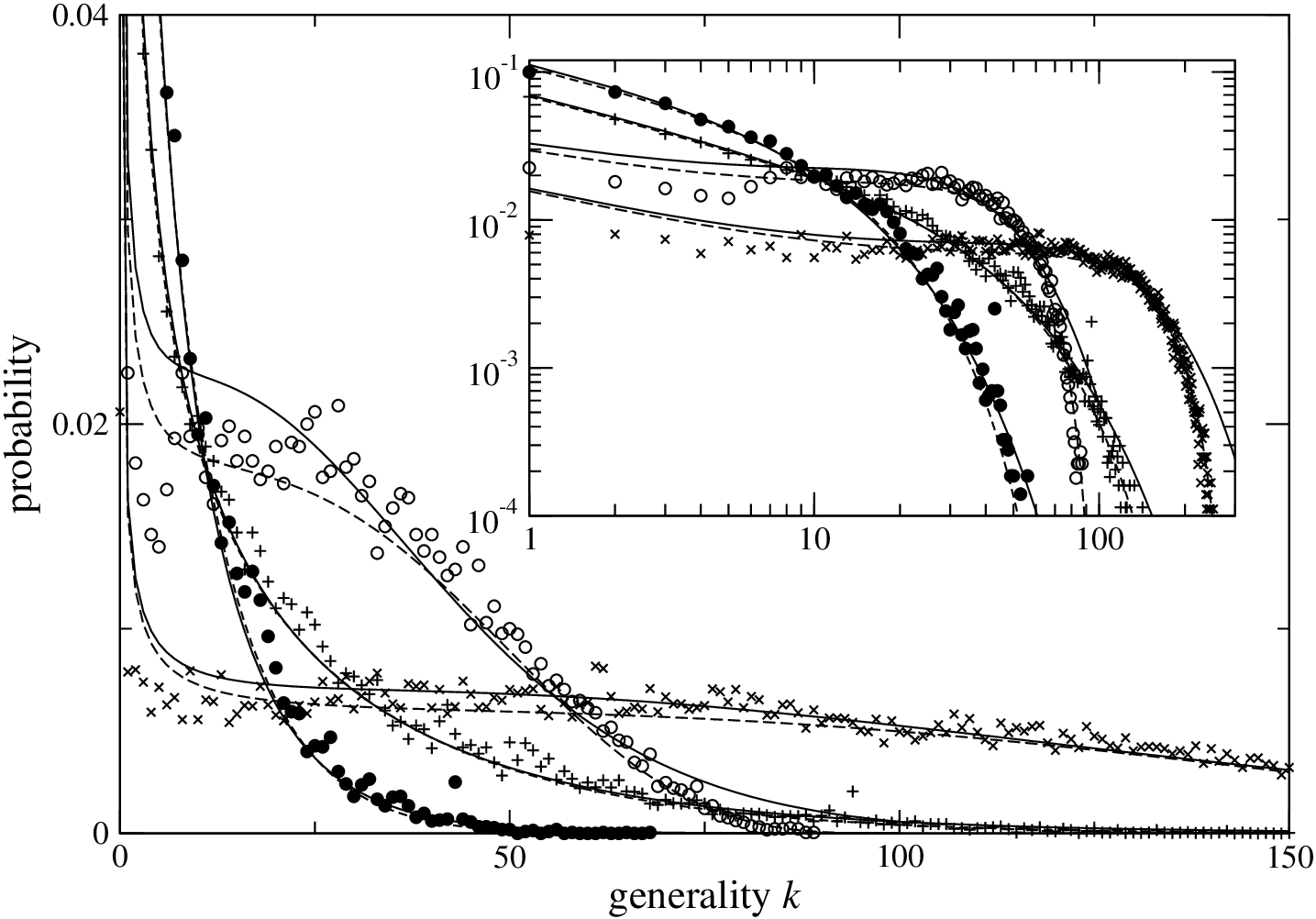}
  \caption{Steady-state generality distributions conditional to fixed
    species number $S$ obtained from simulations with $C_0=0.1$,
    $S=100$ ($\bullet$), $C_0=0.1$, $S=300$ ($+$), $C_0=0.5$, $S=100$
    ($\circ$), and $C_0=0.5$, $S=300$ ($\times$) in comparison with
    the corresponding predictions by Eq.~(\ref{Mgen-fixed-overall})
    (solid) and by directly averaging Eq.~(\ref{gen-dist}) over
    $\Lambda=0..R$ (dashed).  The other parameters were $R=\ln
    10^{20}$, $D=0.005$, $\rho=0.95$, $\kappa=10/R$,
    $\lambda=10^{-3}$, $\beta=0$ (no fitting). For all examples
    $\left<S\right>=190$.  The inset shows the same data on a
    double-logarithmic scale.}
  \label{fig:fixed-S-dists}
\end{figure}

\begin{figure}[tbp]
  \centering
  \includegraphics[width=\imgwidth,keepaspectratio,clip]{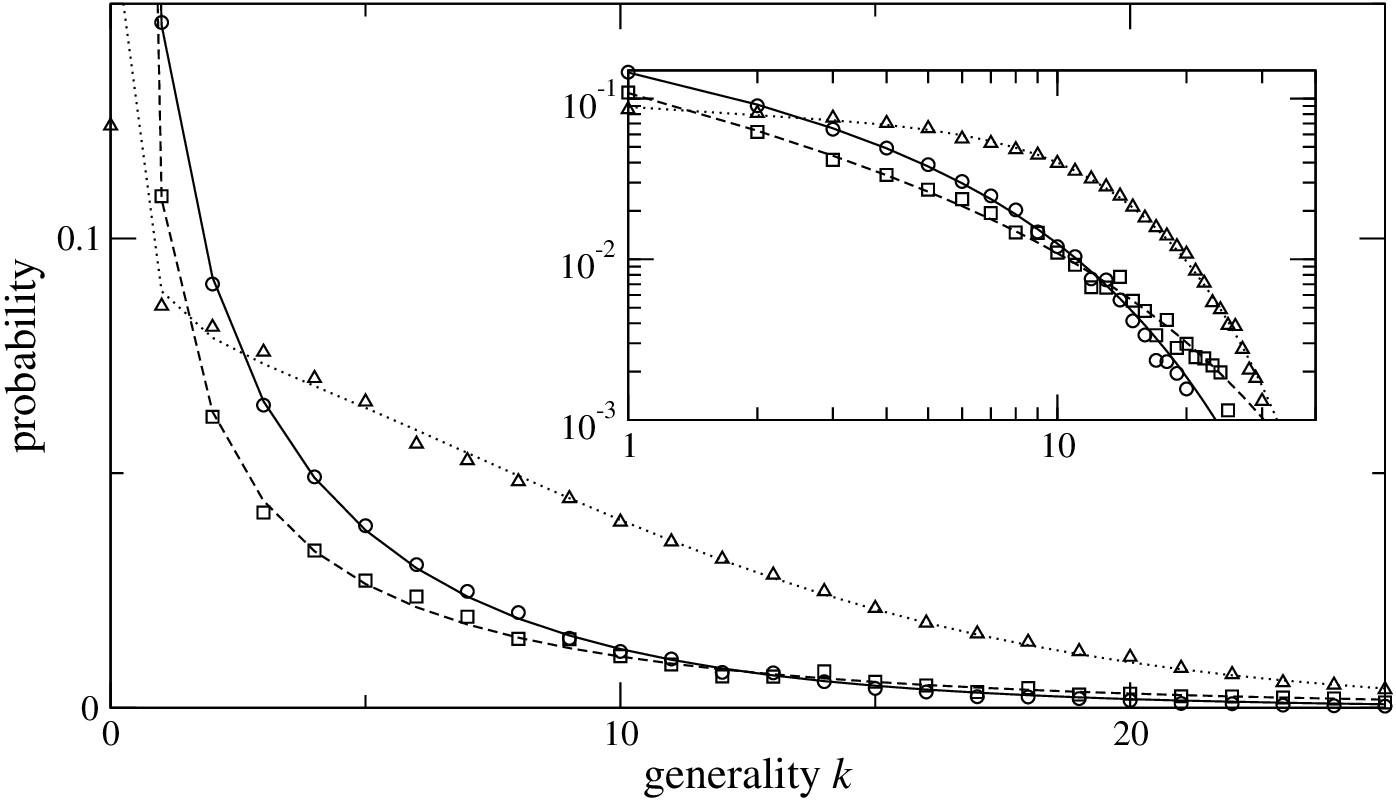}
  \caption{Simulation results for the generality distributions
    conditional to $S=40$ with $C_0=0.1$, $\beta=0$ (squares),
    $C_0=0.1$, $\beta=0.05$ (circles), $C_0=0.3$, $\beta=0.05$
    (triangles) compared to distributions fitted by adjusting the
    parameters $\tilde C$, $\tilde \rho$ and $\tilde \kappa$ in
    Eq.~(\ref{Mgen-fixed-overall}) (dashed, solid, dotted line). The
    other parameters were $R=\ln 10^4$, $D=0.005$, $\rho=0.95$,
    $\kappa=2/R$, $\lambda=10^{-3}$. }
  \label{fig:fitted-dists}
\end{figure}

In order to obtain the overall conditional generality distribution we
go, again, over to the moment-generating function
\begin{align}
  \label{Mgen-fixed}
  M_\text{gen}(s,z|S):=\sum_{k=0}^\infty P_\text{gen}(k,z|S) z^k\approx
  \mathcal{N}\,
  \left(
    \frac{1-\tilde \rho}{1-\tilde \rho z}
  \right)^{\textstyle C_0 \tilde \kappa \Lambda(s)}.
\end{align}
The average of this expression over $s$ for the simple case
$\lambda\to0$ is
\begin{align}
  \label{Mgen-fixed-overall}
  M_\text{gen}(z|S)\approx\frac{\tilde u-1}{\log \tilde u}\quad\text{with }
  \tilde u=  \exp\left(\frac{\vphantom{C}\smash{\tilde C}}{2}\right)\,
    \left(
      \frac{1-\tilde \rho}{1-\tilde \rho\,z}
    \right)^{\textstyle C_0 \tilde \kappa R}.
\end{align}
This result was verified by comparison with a direct numerical
simulations of the model.  Figure~\ref{fig:fixed-S-dists} shows
simulation results in comparison with the predictions of
Eq.~(\ref{Mgen-fixed-overall}) and with the results of numerically
averaging Eq.~(\ref{gen-dist}) directly over $\Lambda=0..R$.  Although
the precision of the approximation Eq.~(\ref{Mgen-fixed-overall})
decreases for increasing $C_0$ and $k$ in comparison with the
prediction using Eq.~(\ref{gen-dist}), it is surprisingly good even
for large values of $C_0$ and $k$.  For large $C_0$ \emph{and} small
$k$ the simulations deviate noticeably also from the prediction using
Eq.~(\ref{gen-dist}), because in this parameter range the effects of
intra-clade consumption, that had here been ignored, become relevant.
Even for smaller $R$, $\kappa R=\mathcal{O}(1)$, and $\beta>0$, where
Eq.~(\ref{Mgen-fixed-overall}) does not make quantitative predictions,
the general form of this expression still seems to be valid.
Figure~\ref{fig:fitted-dists} shows some examples of numerical results
in this regime compared with curves obtained by fitting $\tilde C$,
$\tilde \rho$ and $\tilde \kappa$ in Eq.~(\ref{Mgen-fixed-overall}).
The fitted curves describe the distributions similarly well as the
quantitative predictions above: deviations occur many for very small
and very large $k$.

\subsection{The  vulnerability distribution for fixed $S$}
\label{sec:vulnerability-distribution}

\begin{figure}[tbp]
  \centering
  \includegraphics[width=\imgwidth,keepaspectratio,clip]{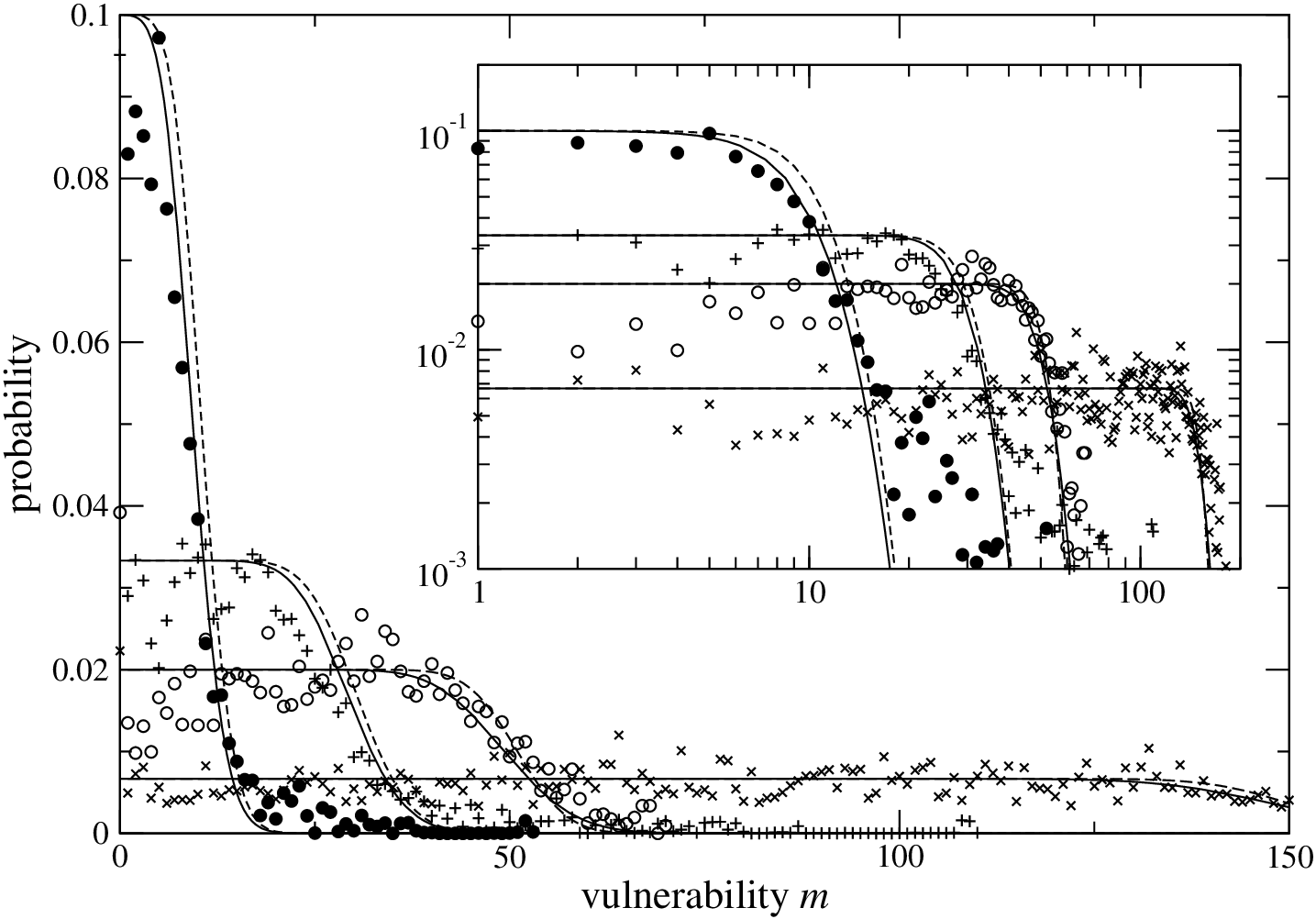}
  \caption{Steady-state vulnerability distributions conditional to fixed
    species number $S$.  Parameters are the same as in
    Fig.~\ref{fig:fixed-S-dists}.  The solid and dashed lines
    correspond to Eqs.~(\ref{vul-dist-int}) and (\ref{vul-dist-sum}),
    respectively.  The inset shows the same data on a
    double-logarithmic scale.}
  \label{fig:fixed-S-vul}
\end{figure}

The distribution of the vulnerability $m$ is most easily computed
directly conditional to fixed $S$: Assume species to be indexed in the
order of increasing $s$ starting with $1$.  For $\lambda\to0$ the
number of possible consumers of species $i$ is then simply $i$.  When
assuming again that resources evolve much faster than their consumers,
the consumers of $i$ are determined by (i) the random connection of
consumers with probability $C_0$ when the resource-clade founder
enteres the food web and (ii) random re-connections with probability
$C_0$ during speciations of resources.  Neither of this processes
introduces correlations in the connectivities within the set of
possible consumers of $i$.  Thus, links are statistically independent
and the vulnerability of $i$ is given by a binomial distribution.
Averaging over the food web yields
\begin{align}
  \label{vul-dist-sum}
  P_\text{vul}(m|S)=\frac{1}{S} \sum_{i=m}^{S} \binom{i}{m} C_0^m (1-C_0)^{i-m}.
\end{align}
This is exactly the expression that
\cite{camacho02:_analytic_food_webs} obtained in their analysis of the
niche model.  Following their observation that in the limit of large
$S$ with constant $Z_0=C_0 S$ and $i=\mathcal{O}(S)$ the binomial
distribution can be approximated as Poisson and the sum by an
integral, one obtains
\begin{align}
  \label{vul-dist-int}
  P_\text{vul}(m|S)=\frac{1}{Z_0}\int_0^{Z_0} \frac{t^m\,e^{-t}}{m!}  dt.
\end{align}
As is shown in Fig.~\ref{fig:fixed-S-vul}, this result predicts the
vulnerability distribution similarly well as
Eq.~(\ref{Mgen-fixed-overall}) the generality distribution.

Note that the Poisson distribution entering Eq.~(\ref{vul-dist-int})
is the special case $P(t/B,B;n)$, $B\to0$ of the general distribution
$P(A,B;n)$ entering Eq.~(\ref{compare-dists}).  Thus, the
integral~(\ref{vul-dist-int}) is also a limiting case of the general
form Eq.~(\ref{Mgen-fixed-overall}).  In the case of generality
distributions, however, $B$ is typically close to one.

\section{Comparison with other topological food-web models}
\label{sec:other-models}

\subsection{Comparison with  the cascade model}
\label{sec:cascade}

The main idea upon which the cascade model is based, random
connections restricted by a trophic hierarchy, is retained in the
speciation model, albeit refined in several ways.  The cascade model
is recovered from the speciation model in the limit of no loops
($\lambda=0$), and no speciations\footnote{Observe that for $r_+\to 0$
  the often encountered combination $-\kappa \ln (1-\rho)$ simplifies
  to $r_1/r_-$.}, i.e. $r_+\to 0$.  Then all species enter the species
pool by adaptations and are independently, randomly connected to their
resources and consumers, just as it was assumed for the consumers
alone in the foregoing section.  However, the limit $r_+\to 0$ does not
describe empirical data particularly well \citep{rossberg05:_web}.
Typical parameter sets for the speciation model have $r_+\approx r_-$
(Tab.~\ref{tab:parameters}).

\subsection{Comparison with the niche model}
\label{sec:niche_model}


\subsubsection{Degree distributions}

It was mentioned already that the distribution of vulnerability in the
niche model is approximately the same as in the speciation model, in
both cases given by Eq.~(\ref{vul-dist-int}).  In the case of the
niche model $Z_0=2 C S=2 Z$ where the targeted connectance $C$ and
the species number $S$ are parameters of the model.  In both cases the
distribution is due to random connections with possible consumers.

For the generality distribution the situation is more complex.  As
the analysis of \cite{camacho02:_analytic_food_webs} showed, it is
for the niche model essentially determined by the distribution of the
``niche width'', i.e., the size of the interval containing the
resources of a species on the niche-parameter scale.
\cite{williams00:_simpl} chose this width for each species as its niche
value $n$ times a random variable $x$ with a beta distribution of the
form
\begin{align}
  \label{beta-dist}
  p_x(x)=b(1-x)^{(b-1)}\approx b\,e^{-b x},
\end{align}
where $b=(1-2\,C)/(2\,C)$ depends on the targeted directed
connectance $C$.  The approximation by an exponential is valid for
$b\gg 1$, i.e. for $C\approx1/(2b) \ll 1$. \cite{williams00:_simpl}
used this particular form for its computational simplicity.  No
ecological arguments to motivate it seem to be known.  Since species
are independently and evenly distributed with density $S$ in the
one-dimensional ``niche space'', the number of species in the niche
interval follows a Poisson distribution with expectation value $S n x$
when $x$ is fixed.  Averaging over all $x$ yields the geometric
distribution
\begin{align}
  \label{niche-generality-distribution}
  P_\text{gen}^{(\text{niche})}(n,k)
  =
  \int_0^{\infty} \frac{(S n x)^k}{k!} e^{-S n x} b\,e^{-b x} dx 
  = 
  \frac{1}{1+n Z_0} \left(
    \frac{n Z_0}{1+n Z_0} \right)^k.
\end{align}


The overall generality distribution is obtained by averaging
Eq.~(\ref{niche-generality-distribution}) over $n$.  The calculation
is simplified by the approximation $k \approx S n x$, i.e.\ 
\begin{align}
  \label{niche-generality-camacho}
  P_\text{gen}^{(\text{niche})}(n,k)\approx \frac{1}{S n}p_x\!\left(\frac{k}{S n}\right)=\frac{1}{n Z_0}\exp\left(-\frac{k}{n Z_0}\right),
\end{align}
which is valid for $n Z_0\gg 1$ [cf.\ 
Eq.~(\ref{niche-generality-distribution})].  This leads to the result of
\cite{camacho02:_analytic_food_webs}
\begin{align}
  \label{niche-overall-generality-camacho}
  P_\text{gen}^{(\text{niche})}(k)=\int_0^1
  P_\text{gen}^{(\text{niche})}(n,k) dn\approx
  \frac{1}{Z_0}E_1(-\frac{k}{Z_0})
\end{align}
with $E_1(x):=\int_x^\infty t^{-1}\exp(-t)dt$ denoting the exponential
integral function.  \cite{camacho02:_robus_patter_food_web_struc}
concluded that the distribution of the scaled generality $k/(2 Z)$ or,
for single instances of food webs more appropriate, its cumulative
distribution, should have the universal form
\begin{align}
  \label{universal-cumulative-generality}
  P\left(\frac{k}{2 Z} \ge x\right)=\int_x^{\infty}
  E_1(x')dx'=\exp(-x)-x E_1(x),
\end{align}
and verified this impressively by a comparison with empirical data.

\begin{figure}[tbp]
  \centering
  \includegraphics[width=0.7\imgwidth,keepaspectratio,clip]{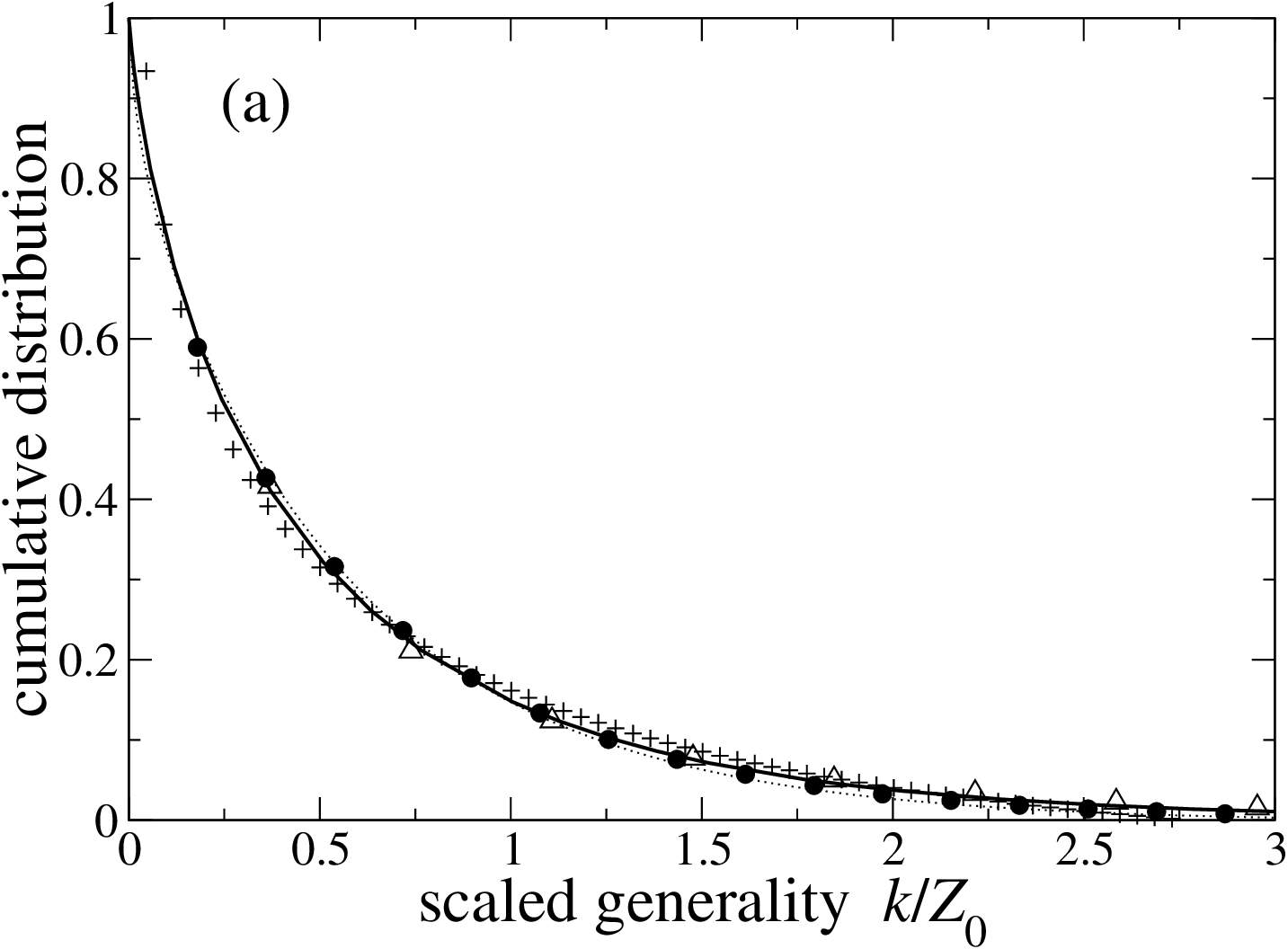}
  \includegraphics[width=0.7\imgwidth,keepaspectratio,clip]{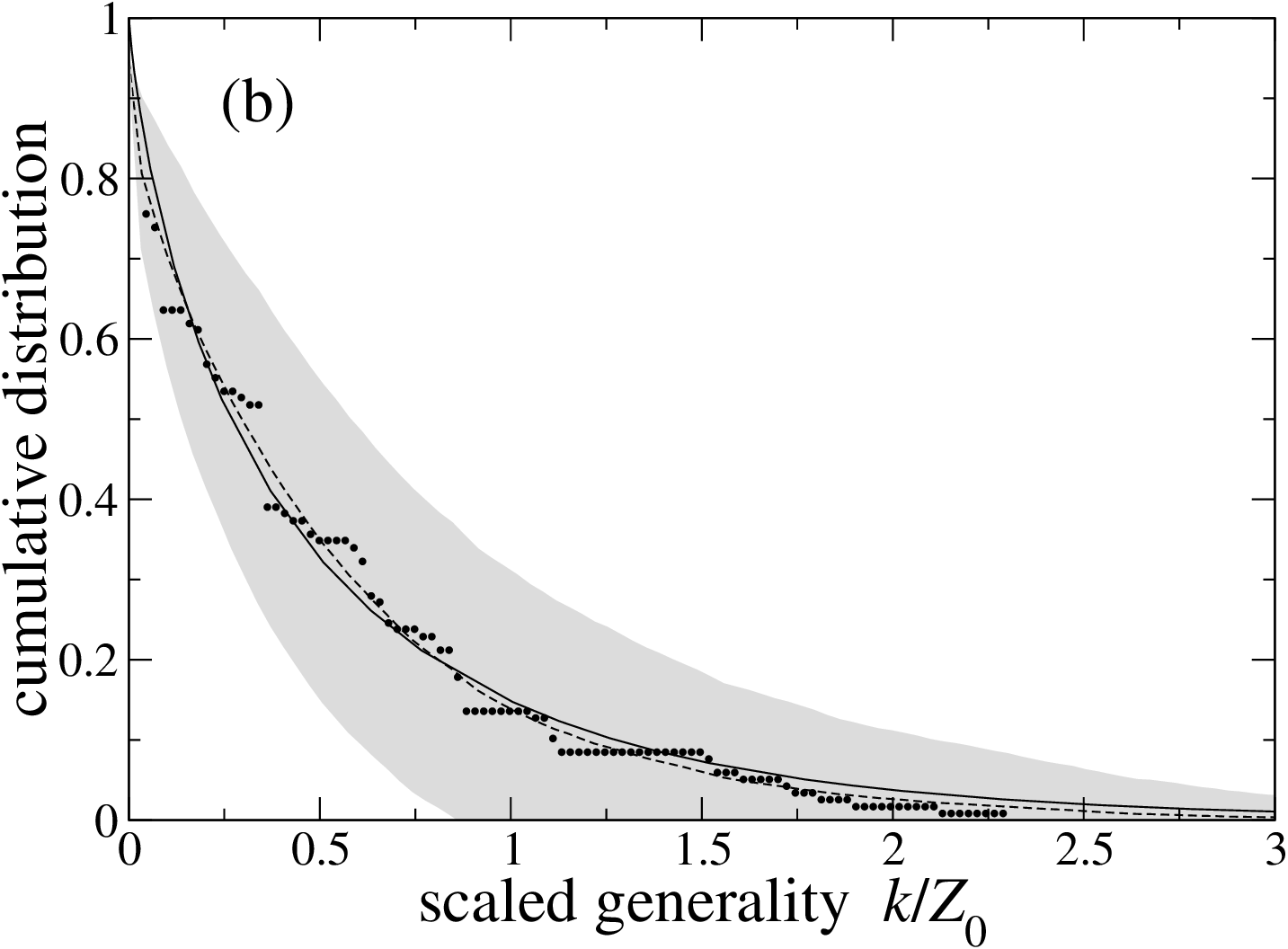}
  \caption{Comparison of niche-model and speciation-model predictions
    for the cumulative generality distribution.  (a) The
    approximation~(\ref{Mgen-fixed-overall}) for the speciation model
    with $C_0\tilde \kappa R=0.2$, $\tilde \rho=0.75$
    (triangles), $C_0\tilde \kappa R=1.5$, $\tilde \rho=0.75$
    (circles), $C_0\tilde \kappa R=0.2$, $\tilde \rho=0.98$
    (plus), $C_0\tilde \kappa R=1.5$, $\tilde \rho=0.98$ (dotted line)
    in comparison with the
    approximation~(\ref{universal-cumulative-generality}) for the niche
    model (solid line). (b) The empirical distribution for Little Rock
    Lake~\cite{martinez91:_artif_attr} (dots), the speciation model
    prediction from numerical simulations (dashed, shaded area is the
    1-$\sigma$ range of fluctuations before scaling), and again the
    approximation~(\ref{universal-cumulative-generality}) for the niche
    model (solid line).  All distributions have the point $k=0$
    removed and are scaled and normalized to minimize mean-square
    deviations from Eq.~(\ref{universal-cumulative-generality}).}
  \label{fig:curve-fitting}
\end{figure}

In order to see if this observed regularity is reproduced also by the
speciation model, cumulative distribution functions for the speciation
mode obtained from Eq.~(\ref{Mgen-fixed-overall}) were compared with
Eq.~(\ref{universal-cumulative-generality}).  The value for $k=0$ was
excluded from the comparison because (i) in many empirical food-webs
the lowest trophic level ($k=0$) is only poorly resolved and (ii) the
approximation (\ref{niche-overall-generality-camacho}) is undefined at
$k=0$ and Eq.(\ref{Mgen-fixed-overall}) is not accurate at this point
either.  The scaling factor $Z_0^{-1}$ for the generality and the
correction $\mathcal{\tilde N}$ of the normalization constant were
therefore determined directly by transforming the cumulative
speciation-model distributions to $\mathcal{\tilde
  N}\sum_{k'=k}^\infty\,P_\text{gen}(k'/Z_0|S)$ such as to minimize
the mean-least-square deviation
from~(\ref{universal-cumulative-generality}) for $k\ge 1$.  These
curves match Eq.~(\ref{universal-cumulative-generality}) surprisingly
well over a wide parameter range (Fig.~\ref{fig:curve-fitting}a).  The
empirical data is described well by both distributions
(Fig.~\ref{fig:curve-fitting}b).

To understand the reason for this apparent scaling law of
speciation-model food webs, consider the speciation-model generality
distribution~(\ref{Mgen-fixed-overall}) conditional to $k\ge 1$ in the
limit of low connectance $C_0,\tilde C\to 0$ (now at fixed $S$),
i.e.\ the distribution with the moment generating function
\begin{align}
  \label{low-C-limit}
  \lim_{C_0,\tilde
      C\to0}\frac{M_\text{gen}(z|S)-M_\text{gen}(0|S)}{1-M_\text{gen}(0|S)}=\frac{\ln(1-\tilde\rho
      z)}{\ln (1-\tilde\rho)}.
\end{align}
This is easily seen to be the distribution of resources-clade sizes
[cf. Eq.~(\ref{clade-size-distribution})]
\begin{align}
  \label{lower-animal-spec}
  \frac{{\tilde \rho}^k}{k\ln(1-\tilde \rho)}.
\end{align}
In this limit of low connectance most species belong to the lowest
trophic level, only a few heterotrophs remain, and the percolation of
the network is lost.  Therefore, this limit does not correspond to the
general situation encountered in the field.  But the
approximate form of the log-series
distribution~(\ref{lower-animal-spec}) is retained also for more
complex networks.  For values of $\tilde \rho\approx 0.8$, this
distribution has a shape quite similar to the exponential integral
distribution Eq.~(\ref{niche-generality-camacho}).  When going over to
cumulative distributions, the fit looks even better.  Thus, the
observed generality distributions can be interpreted mechanistically
in terms of the steady-state distributions of evolutionary clade
sizes, corrected for fixing $S$ and trophic link breaking.  This also
suggests that the ``scaling'' distribution~(\ref{lower-animal-spec})
$\sim k^{-1} \exp(-k \ln \tilde \rho)$ or the more accurate
result~(\ref{Mgen-fixed-overall}) would rather be the adequate
functional forms than the exponential integral
function~(\ref{niche-generality-camacho}).

In spite of the similarities of the overall generality and
vulnerability distributions, there are marked differences in the
detailed predictions of the two models.  Consider, for example, the
generality distribution for species near the lower end of the trophic
cascade, i.e., species with $\kappa \Lambda(s)\ll 1$ in the speciation
model and $n\ll 1$ in the niche model, that have at least one resource
species ($k\ge 1$).  For the speciation model
Eqs.~(\ref{compare-dists}) and (\ref{ABdist}) lead again to the
clade-size distribution~(\ref{lower-animal-spec}), while for the niche
model Eq.~(\ref{niche-generality-distribution}) predicts
\begin{align}
  \label{lower-animal-niche}
  (1-n Z_0)\,(n Z_0)^{k-1}. 
\end{align}
Thus, for the niche model it is very probable that such a species has
exactly one resource, whereas for the speciation model larger
generalities can also be expected.  An empirical test should be
capable of distinguishing these two predictions.

\subsubsection{Intervality}

A major distinction of the niche model from the cascade model is the
intervality it enforces upon the diets of consumers.  While the degree
of intervality obtained with the cascade model is typically too small
compared with empirical data \citep{cohen90:_commun_food_webs}, it is
too large for the niche model \citep{cattin04:_phylog}.  Under certain
conditions the speciation model can also produce a high degree of
intervality.  Consider some arbitrary ordering of clades, for example
by the speed parameter of the founder species, and an ordering of the
species within each clade given by a traversal of the evolutionary
tree\footnote{For example, the order given by the recursive algorithm
  \texttt{list(\textit{A})}
  defined as \texttt{\\
    \hspace*{4ex}1. if \textit{A} has not become extinct\\
    \hspace*{12ex}print \textit{A};\\
    \hspace*{4ex}2. for all direct descendants \textit{B} of
    \textit{A}
    in order of appearance\\
    \hspace*{12ex}list(\textit{B});
  }\\
  starting with \texttt{list(}clade founder\texttt{)}.
}. %
For this ordering (which differs from an ordering by $s$) diets will
form contiguous sets when (i) the average number of resources
clades~(\ref{clade-in-diet}) is low, i.e., when most consumers have
either one or no resource clade, and (ii) the probability that
resources break out of a resource clade during the clade's lifetime is
low.  Then the set of a consumer's resources is usually simply the
non-extinct part of an evolutionary subtree.  The probability of
resource beak-out is small when $\beta\times\text{(resource clade
  size)}\times\text{(clade lifetime in generations)}$ is small, which,
by arguments analogous to those used in Section~\ref{sec:clades}, the
case when 
\begin{align}
  \label{breakout}
  \beta \rho^*/(1-\rho^*)\ll 1.
\end{align}
For typical model parameters we find that these two conditions are
satisfied to some extent but not too well (Tab.~\ref{tab:parameters}),
in accordance with expectations.  Correspondingly, the degree of
intervality $D_\text{diet}$ \citep{cattin04:_phylog} of empirical data
is reproduced well by the model \citep{rossberg05:_web}.


We conclude with \cite{cattin04:_phylog} that the larger-than-random
intervality observed in food webs may not so much result from a low
dimensionality of the niche space, as has been proposed
\citep{cohen78:_food_webs_niche_space}, but rather reflects the
importance of the phylogenetic history for the food-web structure.

\subsection{Comparison with the nested hierarchy model}
\label{sec:hierarchy}

Just as for the niche model, the generality distribution for the
nested hierarchy model is imposed ``by hand'' by specifying the
distribution~(\ref{beta-dist}) and setting $k\approx S n x$.  But the
structure of the set of resources is determined by a more complex
algorithm that has been designed in such a way that consumers and
resources form groups ($\approx$ clades), and consumers and resources
from the same groups share resources and consumers, respectively.  The
algorithm is intended to mimic a structure that would result from a
phylogenetic evolution of the web, without explicitly modeling this
evolution.  The speciation model achieves a similar effect by
explicitly modeling the evolutionary dynamics.

\section{Variants of the speciation model}
\label{sec:variants}

Modeling complex ecological systems often requires difficult decisions
with regards to which kinds of effects ought to be incorporated into a
model and which can be ignored.  Here, two variants of the speciation
model are shortly discussed that include aspects of the real system
that had been left out in the original model.  For both variants, the
analytic results derived in the previous sections remain valid without
change.

\subsection{A variant with  asymmetric link persistence}
\label{sec:asymmetric}

In the analysis above it was assumed that consumer-resource links are
statistically independent of the phylogenetic history of the
consumers.  If this assumption is valid, one may as well modify the
model such as to choose all resources of a descendant species at
random after its speciation, without affecting the analytic results
obtained above.  More generally, one might incorporate an asymmetry in
the persistence (or reconnection probability) of links between
consumers and resources in the following way:

In the original form of the model, the connectivity of the descendant
species was (randomly) re-assigned for a fraction $\beta$ of all
possible trophic links.  In the asymmetric variant of the model, the
connectivity from the descendant species to its consumers is
re-assigned for a fraction $\beta_\text{c}$ of all possible consumers,
and the connectivity to resources is re-assigned for a fraction
$\beta_\text{r}$ of all possible resources, with
$\beta_\text{c}\ne\beta_\text{r}$ in general.

In fact, there is no ecological reason to expect
$\beta_\text{c}=\beta_\text{r}$.  A large difference between the
values of $\beta_\text{c}$ and $\beta_\text{r}$ such as considered
above ($\beta_\text{c}=\beta \ll \beta_\text{r}=1$) could be
understood from the assumption that in the competition between
related species their sets of resources are much more important than
their sets of consumers: In order to avoid competitive exclusion,
related species need drastically different sets of resources
($\beta_\text{c}=1$), while there is only little evolutionary pressure
for a descendant species to have a different set of consumers than its
predecessor ($\beta_\text{c}\ll1 $).

However, one might also argue that by the direct resource-consumer
interaction alone.  Then one could expect it to be advantageous for a
descendant species to evade its predecessors consumers (large
$\beta_\text{c}$), while maintaining its resources (small
$\beta_\text{r}$).  This would lead to the reverse relation between
$\beta_\text{c}$ and $\beta_\text{r}$.  An empirical test to establish
which of these two mechanisms is more relevant might be possible.

\subsection{A variant with quantitative link strength}
\label{sec:quantitative-links}

Topological food-web models are often criticised for ignoring the fact
that the link strength in food webs, instead of being either $1$ or
$0$, is in reality a continuous quantity
\citep{berlow04:_interac_strength}.  There is a simple way to
incorporate continuously varying link strengths in the speciation
model without affecting its statistical properties.

Instead of assigning to each possible trophic link a connectivity of
either $0$ and $1$, quantify the strength of each possible link by are
real number between $0$ and $1$.  Where the connectivity was copied
during speciations in the original model, the links strength is copied
now.  Where the connectivity was set to $1$ with probability $C_0$ and
to $0$ otherwise, set the link strength to an appropriately
distributed random number between $0$ and $1$ now.  For a
characterization of the resulting food webs in terms of topological
food-web statistics, count each link with strength larger than some
threshold as present, and all other links as absent.  That is, the
thresholding of the link strength is just delayed to the time of the
characterization.
While this modification is straightforward for the speciation model,
modifications of other topological models to postpone the thresholding
of link strength might be possible, if at all, only at the price of
increasing the model complexity.

Of course, an evolution where the link strength either does not changes
at all or is reset to a completely new random value is quite
artificial.  More natural it would be to vary the link strength by a
small random amount at each evolutionary step.  In such a model, link
breaking and reconnecting events relative to some threshold $(1-C_0)$
would be correlated.  They would be concentrated at certain pairs of
consumer and resources clades with link strength near the threshold.
Further studies are required to understand what effect this would have
on the overall network structure.

\section{Discussion and Outlook}
\label{sec:conclusion}

Besides improving the general understanding of the properties of the
speciation model and their dependence on model parameters, a purpose
of this work was also to show that the speciation model integrates the
underlying ideas from previous, simpler models (see
Section~\ref{sec:other-models}).
The speciation model retains the trophic ordering of the cascade
model.  In fact, it contains the cascade model as a special case.  By
the interplay of speciations, extinctions, and adaptations of new
species to the habitat, the speciation model reproduces three key
features of the niche model and the nested hierarchy model at the same
time: (1) the empirical distributions of generality, which in the
niche model and similarly in the nested hierarchy model are obtained
only by a special, ecologically unmotivated choice of the niche-width
distributions; (2) intervality, to the degree that is actually
observed \citep{cattin04:_phylog, rossberg05:_web}; (3) the
organization of resources into groups of related species that share
consumers and \textit{vice versa}.
This unifying character of the speciation model is probably the main
reason for its high accuracy in reproducing empirical data
\citep{rossberg05:_web}.


The observed broad, log-series-like generality distributions have been
traced back to, among others, a condition $1-\rho \ll 1$.  This means
that the rate constant for speciations $r_+$ is numerically close to
the rate constant for extinctions $r_-$.  For any phylogenetically
closed system, a steady state always requires that extinction rates
and speciation rates are equal, independent of the statistical details
of the branching pattern.  For the half-open system considered
here, $1-\rho \ll 1$ implies that the contributions from foreign
adaptations to the species pool are small compared to the contribution
from speciations.  In fact, $1-\rho$ directly equals the fraction of
species in the food web that have entered by foreign adaptations.
However, in order to obtain broad, left-skewed generality
distributions, the independence of the speciation and extinction
probability of a species from the actual size of its clade is also
important.  If, instead, large clades would notably favor extinctions
and small clades speciations, clade size distributions would be
dominated by a ``typical'' clade size, which would, in the model, also
lead to a narrower generality distribution.  In an analysis of
paleontological time series \citet{raup91:_phaner_kill} applied a
model for the size of genera identical to the model used here for the
dynamics of clade sizes
[Eqs.~(\ref{probability_balance})-(\ref{jnnp})].  While, on the
average, this model (with $\rho=0.996$) reproduced the data well, the
scatter in the paleontological data was larger than in the model.
\citeauthor{raup91:_phaner_kill} could explain this observation by
assuming that the overall evolution rate varies over time.  Since such
a variation can also be described by a (random) nonlinear
transformation of the time axis, it does not affect statistics that
refer only to a particular moment in time, such as food-web
structures.  Thus our assumption of a simple birth/death process is
supported by paleontological observations.

As a direct consequence of this birth/death process, a
characterization of food webs in terms of ``clades'' has been derived.
Table~\ref{tab:parameters} lists expectation values for characteristic
quantities corresponding to some empirical food webs.  It might be
interesting to compare these results with the taxonomic structure of
the actual empirical webs or the model dynamics with paleontological
records.

In Section~\ref{sec:distributions} it was shown that a correlation
between the evolution rates and the trophic height leads to the
observed asymmetry between generality and vulnerability distributions.
However, in the present model this requires evolution rates spanning
an unrealistically large range of about 20 orders of magnitude.  We
are currently evaluating a variant of the speciation model that
achieves a similar effect without any differences in evolution rates
by making not directly the trophic links hereditary but the properties
of species determining link strengths.  An asymmetry of the heredity
between species-as-consumers and species-as-resources leads
effectively to an asymmetry of the link persistence as described in
Section~\ref{sec:asymmetric} above.  Numerical results with the new
model are promising, but analytically we understand it only in so far
as it can be approximated by the speciation model, so that the analysis
presented here remains valid.  Details regarding the new model will be
reported elsewhere.

Our findings indicate that a food web's population dynamical
stability and persistence are not as important determinants of its
structure as is sometimes assumed.  From a technical point of view,
this is good news.  It appears possible to obtain natural food-web
structures without time-consuming population dynamical simulations.
These food webs could then be investigated also with respect to the
question how their structure affects population dynamical stability.

In the course of this work, analytic approximations for several
empirically testable predictions of the speciation model could be
obtained.  These include the average clade size $\left<n\right>$, the
number of clades $\left<c\right>$ in the web, the age of clades in
generations (speciation times) $-\ln(1-\rho)$, the average number of
resource clades Eq.~(33), and the generality distribution of consumers
at low trophic levels Eq.~(53).
A careful comparison of the models discussed here and other food-web
models with existing empirical data and new results from ongoing
efforts in the field will reveal discrepancies and, hopefully, suggest
new ideas to bringing us another step closer to understanding this
fascinating aspect of life on earth.

\section{Acknowledgements}
\label{sec:thanks}

The authors acknowledge generous support by The 21st Century COE
Program ``Bio-Eco Environmental Risk Management'' of the Ministry of
Education, Culture, Sports, Science and Technology of Japan.

\appendix

\section{A family of distribution functions encountered in the
  analysis of the speciation model}
\label{sec:general}

\renewcommand{\theequation}{A.\arabic{equation}}
\setcounter{equation}{0}

The analysis of the steady-state of a simple model of evolutionary
dynamics (Sec.~\ref{sec:S}) naturally leads to probability
distributions $p_n$ for species number $n$ with a cumulant generating
function
\begin{align}
  \label{generalK}
  \ln\sum_n p_n z^n=K_{A,B}(z)=A \ln
  \left(
    \frac{1-B}{1-B z}
  \right),
\end{align}
where $0< A$, $0<B<1$.

From this, the mean
\begin{align}
  \label{meanAB}
  \left<
    n
  \right>=\left.
    \frac{dK_{A,B}(e^u)}{du}
  \right|_{u=0}=\frac{A\,B}{1-B},
\end{align}
and variance
\begin{align}
  \label{varAB}
  \var n =\left.
    \frac{d^2K_{A,B}(e^u)}{du^2}
  \right|_{u=0}=\frac{A\,B}{(1-B)^2}
\end{align}
can be calculated directly.  The ratio $(\var n)/\! \left< n \right>$
is $(1-B)^{-1}$ times larger than for  Poisson
distributions.

The distribution function itself is given by
\begin{align}
  \label{ABdist}
  p_n=P(A,B;n):=\,\frac{(1-B)^A \, B^n \, \Gamma(A+n)}{\Gamma(A)\, \Gamma(1+n)}.
\end{align}
This implies that the ratio of consecutive probabilities is
\begin{align}
  \label{pk1pk}
  \frac{p_{n+1}}{p_{n}}=\frac{B\,(A+n)}{1+n}.
\end{align}
In particular, the most probable value is $n=0$ whenever $A\,B<1$.
Since $B<1$, this is always the case when $A\le 1$.  For $A=1$ one gets
exactly a geometric distribution
\begin{align}
  \label{unitA}
  p_n=(1-B)\,B^n,
\end{align}
and for small $A$ Eq.~(\ref{ABdist}) simplifies to the log-series
distribution
\begin{align}
  \label{smallA}
  p_n=
  \left\{
    \begin{matrix}
      1+A\,\ln(1-B) +\mathcal{O}(A^2) & \text{for $n=0$},\\
      \displaystyle\frac{A B^n}{n}+\mathcal{O}(A^2) & \text{otherwise}.
  \end{matrix}
\right.
\end{align}
For small $B$ a Poisson distribution is obtained:  With fixed $AB$,
\begin{align}
  \label{smallB}
  p_n=\frac{(AB)^n}{n!}\,e^{-AB}+\mathcal{O}(B)
\end{align}
uniformly in $n$.  Finally, when $A B \gg 1$ the distribution $p_n$
can be approximated by a Gaussian with mean and variance given by
Eqs.~(\ref{meanAB},\ref{varAB}).

\begin{nowordcount}
\bibliographystyle{elsart-harv} 
\bibliography{/home/axel/bib/bibview}
\end{nowordcount}

\end{document}

